\pdfoutput=1  

\def\mg{R_\mathrm{g}}
\def\ag{a_\mathrm{g}}

\def\virg#1{\textquoteleft #1\textquoteright}

\def\su{\sin u}
\def\cu{\cos u}

\def\bds#1{\mathbf{#1}}
\def\acap{\\ \nonumber \\}

\def\rfr#1{Equation\,(\ref{#1})}
\def\rfrs#1#2{Equations\,(\ref{#1})--(\ref{#2})}

\def\Rfrs#1#2{Equations\,(\ref{#1})--(\ref{#2})}
\def\derp#1#2{\frac{\partial{#1}}{\partial{#2}}}

\def\dert#1#2{\frac{{{\mathit{d}}}{#1}}{{{\mathit{d}}}{#2}}} 
 
\def\eqi{\begin{equation}}
\def\eqf{\end{equation}}
\def\eqia{\begin{eqnarray}}
\def\eqfa{\end{eqnarray}}
\def\rp#1#2{\frac{#1}{#2}}
\def\lb#1{\label{#1}}

\def\bds#1{\boldsymbol{#1}}
\def\nk{n_\mathrm{K}}

\def\ton#1{\left(#1\right)}
\def\qua#1{\left[#1\right]}
\def\grf#1{\left\{#1\right\}}

\documentclass[a4paper, 11pt]{article}

\usepackage[english]{babel}
\usepackage[LGR,T1]{fontenc}
\usepackage[utf8]{inputenc}
\usepackage{w-greek}
\usepackage[nointegrals]{wasysym}
\usepackage{authblk}
\usepackage{abstract}
\usepackage{amsmath}
\usepackage{amsthm}
\usepackage{amscd}
\usepackage{amssymb}
\usepackage{dsfont}
\usepackage{xcolor}
\usepackage{graphicx}
\usepackage{epsfig}
\usepackage{xr-hyper}
\usepackage[pdfstartview=XYZ,
bookmarks=true,
colorlinks=true,
linkcolor=blue,
urlcolor=blue,
citecolor=blue,
pdftex,
bookmarks=true,
linktocpage=true, 
hyperindex=true
]{hyperref}
\usepackage{starfont}
\usepackage{txfonts}
\usepackage[mathlines]{lineno}
\usepackage{tabularx}
\usepackage{booktabs}
\usepackage{fontawesome5}

\usepackage[round]{natbib} 
\newcommand{\appendixsection}[1]{%
  \refstepcounter{section}%
  \section*{Appendix \thesection #1}%
  \addcontentsline{toc}{section}{Appendix \thesection: #1}%
}

\allowdisplaybreaks[1]

\makeatletter
\DeclareRobustCommand\ref{%
    \@ifstar\@refstar\T@ref
  }%
  \DeclareRobustCommand\pageref{%
    \@ifstar\@pagerefstar\T@pageref
  }%
 \makeatother

\DeclareSymbolFont{greekletters}{LGR}{\familydefault}{m}{n}
\DeclareMathSymbol{\qA}{\mathord}{greekletters}{65}
\DeclareMathSymbol{\qB}{\mathord}{greekletters}{66}
\DeclareMathSymbol{\qG}{\mathord}{greekletters}{71}
\DeclareMathSymbol{\qD}{\mathord}{greekletters}{68}
\DeclareMathSymbol{\qE}{\mathord}{greekletters}{69}
\DeclareMathSymbol{\qZ}{\mathord}{greekletters}{90}
\DeclareMathSymbol{\qEt}{\mathord}{greekletters}{72}
\DeclareMathSymbol{\qTh}{\mathord}{greekletters}{74}
\DeclareMathSymbol{\qI}{\mathord}{greekletters}{73}
\DeclareMathSymbol{\qK}{\mathord}{greekletters}{75}
\DeclareMathSymbol{\qL}{\mathord}{greekletters}{76}
\DeclareMathSymbol{\qM}{\mathord}{greekletters}{77}
\DeclareMathSymbol{\qN}{\mathord}{greekletters}{78}
\DeclareMathSymbol{\qX}{\mathord}{greekletters}{88}
\DeclareMathSymbol{\qO}{\mathord}{greekletters}{79}
\DeclareMathSymbol{\qP}{\mathord}{greekletters}{80}
\DeclareMathSymbol{\qR}{\mathord}{greekletters}{82}
\DeclareMathSymbol{\qS}{\mathord}{greekletters}{83}
\DeclareMathSymbol{\qT}{\mathord}{greekletters}{84}
\DeclareMathSymbol{\qU}{\mathord}{greekletters}{85}
\DeclareMathSymbol{\qPh}{\mathord}{greekletters}{70}
\DeclareMathSymbol{\qCh}{\mathord}{greekletters}{81}
\DeclareMathSymbol{\qPs}{\mathord}{greekletters}{89}
\DeclareMathSymbol{\qOm}{\mathord}{greekletters}{87}
\DeclareMathSymbol{\qa}{\mathord}{greekletters}{97}
\DeclareMathSymbol{\qb}{\mathord}{greekletters}{98}
\DeclareMathSymbol{\qg}{\mathord}{greekletters}{103}
\DeclareMathSymbol{\qd}{\mathord}{greekletters}{100}
\DeclareMathSymbol{\qe}{\mathord}{greekletters}{101}
\DeclareMathSymbol{\qz}{\mathord}{greekletters}{122}
\DeclareMathSymbol{\qet}{\mathord}{greekletters}{104}
\DeclareMathSymbol{\qth}{\mathord}{greekletters}{106}
\DeclareMathSymbol{\qi}{\mathord}{greekletters}{105}
\DeclareMathSymbol{\qk}{\mathord}{greekletters}{107}
\DeclareMathSymbol{\ql}{\mathord}{greekletters}{108}
\DeclareMathSymbol{\qm}{\mathord}{greekletters}{109}
\DeclareMathSymbol{\qn}{\mathord}{greekletters}{110}
\DeclareMathSymbol{\qx}{\mathord}{greekletters}{120}
\DeclareMathSymbol{\qo}{\mathord}{greekletters}{111}
\DeclareMathSymbol{\qp}{\mathord}{greekletters}{112}
\DeclareMathSymbol{\qr}{\mathord}{greekletters}{114}
\DeclareMathSymbol{\fs}{\mathord}{greekletters}{99}       
\DeclareMathSymbol{\qs}{\mathord}{greekletters}{115}       
\DeclareMathSymbol{\qt}{\mathord}{greekletters}{116}
\DeclareMathSymbol{\qu}{\mathord}{greekletters}{117}
\DeclareMathSymbol{\qvf}{\mathord}{greekletters}{102}
\DeclareMathSymbol{\qch}{\mathord}{greekletters}{113}
\DeclareMathSymbol{\qps}{\mathord}{greekletters}{121}
\DeclareMathSymbol{\qom}{\mathord}{greekletters}{119}
\DeclareMathSymbol{\sui}{\mathord}{greekletters}{124}
\DeclareMathSymbol{\dig}{\mathord}{greekletters}{147}
\DeclareMathAccent{\uml}{\mathord}{greekletters}{34} 
\DeclareMathAccent{\umld}{\mathord}{greekletters}{35} 
\DeclareMathAccent{\mcr}{\mathord}{greekletters}{31}
\DeclareMathAccent{\aca}{\mathord}{greekletters}{39} 
\DeclareMathAccent{\ga}{\mathord}{greekletters}{96} 
\DeclareMathAccent{\rb}{\mathord}{greekletters}{60} 
\DeclareMathAccent{\smb}{\mathord}{greekletters}{62} 
\DeclareMathAccent{\ca}{\mathord}{greekletters}{126} 
\DeclareMathAccent{\carb}{\mathord}{greekletters}{64} 
\DeclareMathAccent{\casb}{\mathord}{greekletters}{92} 
\DeclareMathAccent{\aarb}{\mathord}{greekletters}{86} 
\DeclareMathAccent{\aasb}{\mathord}{greekletters}{94} 
\DeclareMathAccent{\garb}{\mathord}{greekletters}{67} 
\DeclareMathAccent{\gasb}{\mathord}{greekletters}{95} 

\DeclareTextSymbol{\textquoteleft}{LGR}{28}      
\DeclareTextSymbol{\textquoteright}{LGR}{29}     

\definecolor{orcidlogocol}{rgb}{0.65, 0.807, 0.223}

\newcommand{\orcid}[1]{$\,$\href{https://orcid.org/#1}{\textcolor{orcidlogocol}{\faOrcid}}}

\title{
\textbf{
The measurable impact of the 2pN spin-dependent accelerations on the jet precession of M87$^\ast$
}
}

\author[]{
Lorenzo Iorio\,\orcid{0000-0003-4949-2694}
}

\affil[]{
\href{https://ror.org/01ehyh486}{Ministero dell' Istruzione e del Merito}
\\
Viale Unit\`{a} di Italia 68, I-70125, Bari, Italy \\ email: \href{mailto:lorenzo.iorio@libero.it}{\texttt{lorenzo.iorio@libero.it}}
}

\date{\today}

\providecommand{\keywords}[1]{keywords--- #1}

\begin{document}

\maketitle

\begin{center}
\begin{abstract}
\noindent
Motivated by  recent accurate measurements of disk/jet coprecessions around some galactic supermassive black holes, the accelerations experienced by an uncharged, spinless object in the Kerr metric, written in harmonic coordinates, are analytically calculated up to the formal second post-Newtonian order. To such a level, some new accelerations make their appearance. They are proportional to even and odd powers of the hole's angular momentum. Their counterparts are not known where the primary is a material body.   After expressing them in a coordinate-independent, vector form valid for any orientations of the hole's spin axis in space, their orbital effects are perturbatively worked out in terms of the particle's Keplerian orbital elements. The resulting expressions, averaged over one orbital revolution, are valid for generic shapes and orientations of the orbit. The orbital plane's precession proportional to the first power of the hole's angular momentum and to the reciprocal of the fourth power of the speed of light amounts to about twenty per cent of the corresponding Lense-Thirring effect. The latter is believed to be the cause of the accurately measured disk/jet precessional phenomenology, currently measured to a few per cent accuracy. Although at a lesser extent, also the precession proportional to the second power of the hole's spin and to the reciprocal of the fourth power of the speed of light is measurable. Allowed domains in the parameter space of the jet precession around M87$^\ast$ are displayed.  
\end{abstract}
\end{center}

\keywords{Gravitation\,(661); Celestial mechanics\,(211); Black holes\,(162)}

\section{Introduction}
Recently, the precessions of both the jet and the accretion disk around some supermassive black holes-or megapyknons-lurking in galactic nuclei were measured to a per cent accuracy. In particular, the jet precession around M87$^\ast$ was measured to $\simeq 3\%$  \citep{2023Natur.621..711C}, while the disk/jet coprecession associated with the tidal disruption event AT2020afhd was detected to $\simeq 7\%$ \citep{TDELTCinesi2025}. In both scenarios, the jet and the accretion disk were considered tightly coupled. Furthermore, the latter one was assumed misaligned with respect to the hole's equator. 

It was demonstrated that such precessional phenomenology can be successfully explained in terms of the  Lense-Thirring effect, which is formally of the first post-Newtonian (1pN) order, changing the orbital angular momentum of a fictitious test particle orbiting the black hole-or pyknon-at hand along a circular orbit whose effective radius amounts approximately to ten and a half gravitational radii \citep{2025PhRvD.111d4035I}. 

Given the good experimental accuracies already reached so far and in view of further improvements expected in the next future, the success of the aforementioned effective test particle  scheme demands to accurately investigate all the accelerations that can cause the orbital plane of a fictitious nonspinning and uncharged body to precess in the Kerr spacetime  to higher pN orders than done until now. To this aim and to facilitate the interpretation of the observations avoiding spurious coordinate artifacts, the Kerr metric will be written in harmonic coordinates\footnote{\citet{Fock59} believed that their importance went far beyond simple computational advantages, being of fundamental nature.} \citep{deDon21,Lancz23}, and the geodesic equations of motion of a test particle will be analyzed formally up to the 2pN order. General expressions for the resulting accelerations, expanded up to the fourth power of the hole's spin $S$, will be obtained without restricting to the hole's equatorial plane which, moreover, will be assumed arbitrarily oriented in space. Their orbital effects, averaged over one orbital revolution, will be calculated for general shapes and inclinations of the particle's orbit.

The accelerations will be expressed in a coordinate-independent, vector form, while their orbital effects will be calculated in terms of the standard Keplerian orbital elements in order to make them accessible to the widest possible readership, especially to those acquainted with the widely used formalism of celestial mechanics and dynamical astronomy. These distinctive features  make it difficult to straightforwardly compare the present results with others that can be found in the literature. Indeed, the latter ones were often obtained by using more abstract formalisms.
Last but not least, the order of approximation to which they were obtained is lower than the present one; see, e.g., \citet{2006PhRvD..74l4027L}, \citet{2008PhRvD..78d4012P}, \citet{2013PhRvL.111b1101H} and \citet{2014GReGr..46.1671J}, and references therein.

In addition to some well known accelerations like the 1pN and 2pN mass-only ones, the formally 1pN Lense-Thirring acceleration proportional to\footnote{For a Kerr black hole,  it is of the order of $\mathcal{O}\ton{1/c^3}$.} $S/c^2$, where $c$ is the speed of light in vacuum, the formally\footnote{For a Kerr black hole, they are of the order of $\mathcal{O}\ton{1/c^4}$ and $\mathcal{O}\ton{1/c^8}$, respectively.} \virg{Newtonian} accelerations due to the first two even zonal mass multipoles accounting for the axisymmetric departures from spherically symmetry of the primary and the formally\footnote{For a Kerr black hole, it is of the order of $\mathcal{O}\ton{1/c^6}$.} 1pN quadrupolar acceleration, also other less known-if any-accelerations, proportional to $S^k/c^4,\,k=1,2,3$, will be obtained.

The paper is organized as follows. The transformation of the Kerr metric from the Boyer-Lindquist coordinates to the harmonic ones is performed  in Section~\ref{sec:2}. The resulting accelerations are discussed in Section~\ref{sec:3}. Section~\ref{sec:4} deals with the perturbative calculational scheme to work out their orbital effects. Those of the order of $\mathcal{O}\ton{S/c^4}$ are treated in Section~\ref{sec:5}. Section~\ref{sec:6} is devoted to the consequences of the position-dependent acceleration of the order of $\mathcal{O}\ton{S^2/c^4}$. The calculation of the impact of the acceleration of the order of $\mathcal{O}\ton{S^3/c^4}$ is presented in Section~\ref{sec:7}. 
In Section~\ref{sec:8}, the results obtained in the previous Sections are applied to the M87$^\ast$ scenario by obtaining allowed regions in the two-dimensional spin-orbit radius parameter space suitable for it. Section~\ref{sec:9} summarizes the findings and offers conclusions. Appendix \ref{app:A} contains the details of the transformation of the Kerr metric from the Boyer-Lindquist coordinates to the harmonic ones. Appendix \ref{app:B} describes how to obtain general formulas for any spin-dependent acceleration when expressions valid only for the primary's spin axis aligned with the $z$ axis are available.
\section{The Kerr metric in harmonic coordinates}\lb{sec:2}
The squared spacetime interval $\ton{ds}^2$ of the Kerr metric, written in terms of the Boyer-Lindquist coordinates  $x^0_\mathrm{BL},r_\mathrm{BL},\theta_\mathrm{BL},\phi_\mathrm{BL}$ in a reference frame whose fundamental plane is aligned with the hole's equator, is
\eqi
\ton{ds}^2 = g_{00}^\mathrm{BL}~\ton{dx^0_\mathrm{BL}}^2 + 2 g_{0\phi}^\mathrm{BL}~dx^0_\mathrm{BL}~d\phi_\mathrm{BL} +  g_{rr}^\mathrm{BL}~\ton{dr_\mathrm{BL}}^2 + g_{\theta\theta}^\mathrm{BL}~\ton{d\theta_\mathrm{BL}}^2 +  g_{\phi\phi}^\mathrm{BL}~\ton{d\phi_\mathrm{BL}}^2.\lb{ds}
\eqf
The spacetime metric coefficients $g^\mathrm{BL}_{\qm\qn}$ entering \rfr{ds} can be cast into the form\footnote{In the coefficient of $d\phi_B^2$ of Equation (69) in \citet{2008PhRvD..77j4001H} there is a typo: $\ton{r_B^2 + a^2}$ should be replaced by its squared counterpart $\ton{r_B^2 + a^2}^2$; here, it is meant that $d\phi_B\rightarrow d\phi_\mathrm{BL}$, $r_B\rightarrow r_\mathrm{BL}$ and $a\rightarrow a_\mathrm{g}$ of the present notation.} \citep{2008PhRvD..77j4001H}
\begin{align}
g_{00}^\mathrm{BL} \lb{g00BL}& = \rp{a_\mathrm{g}^2\sin^2\theta_\mathrm{BL} - \Delta}{\rho^2}, \acap
g_{rr}^\mathrm{BL} \lb{g11BL}& = \rp{\rho^2}{\Delta}, \acap
g_{\theta\theta}^\mathrm{BL} \lb{g22BL}& =\rho^2, \acap
g_{\phi\phi}^\mathrm{BL} \lb{g33BL}& = \rp{\sin^2\theta_\mathrm{BL}}{\rho^2}\qua{\ton{r_\mathrm{BL}^2 + a_\mathrm{g}^2}^2 - a_\mathrm{g}^2\Delta \sin^2\theta_\mathrm{BL}}, \acap
g_{0\phi}^\mathrm{BL} \lb{g03BL}& = \rp{\sin^2\theta_\mathrm{BL}}{\rho^2}\qua{a_\mathrm{g}\Delta - a_\mathrm{g}\ton{r_\mathrm{BL}^2 + a_\mathrm{g}^2}}.
\end{align}
In \rfrs{g00BL}{g03BL}, it is
\begin{align}
\rho^2 \lb{rho2} &:= r_\mathrm{BL}^2 + a_\mathrm{g}^2\cos^2\theta_\mathrm{BL}, \acap
\Delta \lb{Delta} &:= r^2_\mathrm{BL} -2\mg r_\mathrm{BL} +a_\mathrm{g}^2.
\end{align}
The dimensional parameters $\mg$ and $a_\mathrm{g}$ entering \rfrs{g00BL}{Delta} are defined as\footnote{The spin parameter $a_\mathrm{g}$ should not be confused with the semimajor axis of the orbit of the test particle which is customarily denoted as $a$ in celestial mechanics.}
\begin{align}
R_\mathrm{g} \lb{mg}&:= \rp{GM}{c^2},\acap
a_\mathrm{g} \lb{a}&:=\rp{S}{Mc},
\end{align}
where $G$ and $c$ are the Newtonian constant of gravitation and the speed of light in a vacuum, while $M$ and $S$ are the hole's mass and spin, respectively. Both \rfr{mg} and \rfr{a} are dimensionally lengths, with \rfr{mg} being nothing else the hole's gravitational radius.

For the purpose of the present paper, it is convenient to express the Kerr metric in terms of the harmonic coordinates, which will be labeled as $x^0,r,\theta,\phi$ in the following.
This task is implemented in Appendix \ref{app:A}.
\section{The acceleration of a test particle from the geodesic equations of motion }\lb{sec:3}
The next step consists of calculating the geodesic equations of motion for the test particle to the desired order of approximation in inverse powers of $c$ in order to obtain its acceleration\footnote{Recovering it to the desired pN order from a Lagrangian would not be as straightforward. Indeed, as pointed out by, e.g., \citet{2006PhRvD..74l4027L}, the equations of motion in harmonic coordinates \citep{1993PhRvD..47.3281K,1993PhRvD..47.4183K,1995PhRvD..52..821K} do not correspond to a conventional Lagrangian \citep{DamDe81a,DamDe81b,1986AIHPA..44..263D}. They can be derived from a generalized Lagrangian depending
on the coordinates, the velocities, as well as a relative acceleration.}. Written in terms of the coordinate time $t=x^0/c$, they are \citep{SoffelHan19}
\eqi
\ddot x^i = -c^2\qua{
\qG^i_{00}
+
2\qG^i_{0j}\rp{\dot x^j}{c}
+
\qG^i_{jk}\rp{\dot x^j}{c}\rp{\dot x^k}{c} 
-
\ton{\qG^0_{00} + 2 \qG^0_{0j}\rp{\dot x^j}{c} + \qG^0_{jk}\rp{\dot x^j}{c}\rp{\dot x^k}{c} }\rp{\dot x^i}{c}
},\, i=1,2,3\lb{eqmo}
\eqf
where 
\begin{align}
\dot x^i &:=\dert{x^i}{t},\, i=1,2,3,\acap
\ddot x^i &:=\rp{d^2 x^i}{dt^2},\, i=1,2,3,
\end{align}
and $\qG^{\qa}_{\qb\qg}$ are the Christoffel symbols of the second kind defined as
\eqi
\qG^{\qa}_{\qb\qg} := \rp{1}{2}g^{\qr\qa}\ton{\derp{g_{\qb\qr}}{x^\qg} +\derp{g_{\qg\qr}}{x^\qb} -\derp{g_{\qb\qg}}{x^\qr} },\,\qa,\qb,\qg = 0,1,2,3.\lb{Chr}
\eqf
In  \rfr{Chr}, $g^{\qm\qn}$ is the inverse metric tensor.

As far as the coordinates $x^i,\,i=1,2,3$ and their time derivatives entering \rfr{eqmo} are concerned, in the present case it is
\begin{align}
x^1 &= r, \acap
x^2 &= \theta, \acap
x^3 &= \phi,\acap
\dot x^1 &=\dot r,\acap
\dot x^2 &=\dot\theta,\acap
\dot x^3 &=\dot\phi,\acap
\ddot x^1 &=\ddot r,\acap
\ddot x^2 &= \ddot\theta,\acap
\ddot x^3 &= \ddot\phi.
\end{align}
It may be useful to recall that, in spherical coordinates, the components of the velocity and acceleration are
\begin{align}
v_r & = \dot r,\acap
v_\theta & = r\dot\theta,\acap
v_\phi & = r\sin\theta\dot\phi
\end{align}
and
\begin{align}
A_r & =  \ddot r -r\ton{\dot\theta^2 + \sin^2\theta\dot\phi^2},\acap
A_\theta & = r\ddot\theta +2\dot r\dot\theta -r\sin\theta\cos\theta\ddot\phi^2,\acap
A_\phi & = r\sin\theta\ddot\phi +2\dot\phi\ton{r\cos\theta\dot\theta + \sin\theta\dot r},
\end{align}
respectively.
The full expressions of the Christoffel symbols for the Kerr metric in harmonic coordinates cannot be explicitly displayed here because of their extreme cumbersomeness.

A computationally efficient way to proceed consists of calculating \rfr{eqmo} starting from suitably approximated expressions of \rfrs{g00}{gthph}  and expanding the result to the desired order in inverse powers of $c$. In the following, the terms of the order of $O\ton{1/c^2}$ and $O\ton{1/c^4}$ will be considered.

Following the aforementioned approach, \rfr{eqmo} yields the acceleration's components
\begin{align}
A_r \lb{Arad}& \simeq -\rp{GM}{r^2} + A_r^{\ton{1/c^2}} + A_r^{\ton{1/c^4}},\acap
A_\theta \lb{Atheta}& \simeq  A_\theta^{\ton{1/c^2}} + A_\theta^{\ton{1/c^4}},\acap
A_\phi \lb{Aphi}& \simeq  A_\phi^{\ton{1/c^2}} + A_\phi^{\ton{1/c^4}}.
\end{align}

As a test of the present results,  the same calculation was repeated to the same order in inverse powers of $c$  by using the  harmonic cartesian coordinates. The starting point was the Kerr metric expressed in terms of the latter ones, as in\footnote{For the parameter $R$ defined in \citet{2014GReGr..46.1671J} by the relation $\ton{x_1^2+x_2^2}/\ton{R^2+\ag^2} +x_3^2/R^2=1$, the solution $R=\sqrt{\qua{x_1^2 + x_2^2 + x_3^2-\ag^2 +\sqrt{4\ag^2 x_3^2+\ton{\ag^2-x_1^2 - x_2^2 - x_3^2}^2}}/2}$ was assumed since it reduces to $\sqrt{x_1^2 + x_2^2 + x_3^2}:=r$ in the limit $\ag\rightarrow 0$.} Equation (15) of \citet{2014GReGr..46.1671J}. It turned out that all the 1pN and 2pN accelerations obtained with the spherical coordinates and displayed in the next Sections agree with those calculated with the cartesian coordinates. Cfr. with the test particle acceleration of Equation (19) of \citet{2014GReGr..46.1671J} which, however, is limited to  terms up to  $1/r^4$ and of second order in $S$.
\subsection{The accelerations of the order of $\mathcal{O}\ton{1/c^2}$}
In \rfrs{Arad}{Aphi}, there are three terms which are formally of the order of $\mathcal{O}\ton{1/c^2}$. They turn out to be the standard Schwarzschild-like acceleration due to the mass $M$ only of the black hole, the Lense-Thirring acceleration of the order of $\mathcal{O}\ton{S/c^2}$ induced by its angular momentum 
\eqi
S = \chi \rp{M^2 G}{c},\, \left|\chi\right|\leq 1 \lb{S}
\eqf 
and the one of the order of $\mathcal{O}\ton{S^2/c^2}$ due to the hole's quadrupole mass moment 
\eqi
Q_2 = -\rp{S^2}{c^2 M}.\lb{Q2}
\eqf
in the special case in which the hole's spin axis $\bds{\hat{k}}$ is aligned with the $z$ axis.
It may be noted that, in view of \rfr{S}, the Lense-Thirring acceleration in the field of a Kerr black hole is, actually, of the order of $\mathcal{O}\ton{1/c^3}$, while the quadrupolar one is  formally Newtonian when $Q_2$ is introduced. 

While the first two accelerations are identical to those arising  in the field of a rotating material body of mass $M$, equatorial radius $R$ and angular momentum $S$ to the first post-Newtonian (1pN) level, i.e. to the order of $\mathcal{O}\ton{1/c^2}$, the latter one corresponds to the purely Newtonian acceleration caused by the body's oblateness $J_2$ after the formal replacement 
\eqi
Q_2 \rightarrow -J_2 MR^2.\lb{GOH}
\eqf

All of them were reobtained from the Kerr metric in harmonic cartesian coordinates as in Equation (15) \citet{2014GReGr..46.1671J}.
\subsection{The accelerations of the order of $\mathcal{O}\ton{1/c^4}$}\lb{acc4}
In \rfrs{Arad}{Aphi}, the acceleration which is formally of the order of $\mathcal{O}\ton{1/c^4}$ consists of the sum of five terms only some of which have directly identifiable equivalents in the case where the primary is a rotating material body. They are the 2pN mass-only acceleration, the acceleration due to the hole's mass multipole of degree $\ell=4$
\eqi
Q_4 = \rp{S^4}{c^4 M^3}\lb{Q4},
\eqf
whose counterpart is the Newtonian acceleration due to the second even zonal harmonic
\eqi
J_4\rightarrow -\rp{Q_4}{MR^4}
\eqf
of the central material source, and a part of the total acceleration of the order of $\mathcal{O}\ton{S^2/c^4}$, the latter being
\begin{align}
A_r^{\ton{S^2/c^4}} \nonumber & = \rp{G S^2}{4 c^4 M r^5}\qua{-24 G M - 9 r \dot r^2 + 3 r^3 \dot\theta^2 + \ton{-56 G M - 27 r \dot r^2 + 9 r^3 \dot\theta^2} \cos 2\theta - \right.\acap
\lb{Ar1}&\left. -48 r^2 \dot r \dot\theta \cos\theta \sin\theta + 3 r^3 \dot\phi^2 \ton{1 + 3 \cos 2\theta} \sin^2\theta},\acap
A_\theta^{\ton{S^2/c^4}} \nonumber & = \rp{G S^2}{8 c^4 M r^5}\grf{-72 r^2 \dot r \dot\theta \cos 2\theta + \qua{-56 G M + 12 r \dot r^2 + 
6 r^3 \ton{-6 \dot\theta^2 + \dot\phi^2}} \sin 2\theta - \right.\acap
\lb{Atheta1}&\left. - 3 r^2 \ton{8 \dot r \dot\theta + r \dot\phi^2 \sin 4\theta}},\acap
A_\phi^{\ton{S^2/c^4}} \lb{Aphi1}& = -\rp{3 G S^2  \dot\phi \sin\theta \ton{\dot r + 3 \dot r \cos 2\theta + 2 r \dot\theta \sin 2\theta}}{c^4 M r^3}.
\end{align}
The aforementioned piece of \rfrs{Ar1}{Aphi1} corresponding  to the 1pN quadrupolar acceleration of the order of $\mathcal{O}\ton{J_2/c^2}$ in the case of an material oblate primary is
\begin{align}
{\bds A}^{\ton{J_2/c^2}} \nonumber &= \rp{GM J_2 R^2}{c^2 r^4}\grf{
\rp{3}{2}\qua{\ton{5 r_k^2 -1}\bds{\hat{r}} -2r_k\bds{\hat{k}} }\ton{v^2 - \rp{4 GM}{r}} - 6\qua{\ton{5 r_k^2 -1}v_r -2 r_k v_k }\bds v -\right.\acap
\lb{A1pNJ2} &\left. - \rp{2GM}{r}\ton{3r_k^2 - 1}\bds{\hat{r}}
}.
\end{align} 
In \rfr{A1pNJ2}, it is
\begin{align}
r_k \lb{q1} &:= \bds{\hat{r}}\bds\cdot\bds{\hat{k}},\acap
v_r \lb{q2} &:= \bds v\bds\cdot\bds{\hat{r}},\acap
v_k \lb{q3} &:= \bds v\bds\cdot\bds{\hat{k}}.
\end{align}
Note that while \rfr{q1} is dimensionless, \rfrs{q2}{q3} have the dimensions of a speed.
The orbital effects due to \rfr{A1pNJ2}, calculated for any orientations of $\bds{\hat{k}}$, can be found in \citet{2024gpno.book.....I}.
Instead, the two  terms of the order of $\mathcal{O}\ton{1/c^4}$ which are proportional to $S$ and $S^3$ do not seem to have any known counterpart where the central object is a material body.  Their orbital effects will be worked out in Sections \ref{sec:5} and \ref{sec:7}. 

About the remaining part of the order of $\mathcal{O}\ton{S^2/c^4}$ which adds to \rfr{A1pNJ2} to form the total acceleration of \rfrs{Ar1}{Aphi1} provided that the replacements of \rfrs{Q2}{GOH} are made, it is less clear if or to which extent it has possible equivalents in the case of a material source. Its orbital consequences will be treated in Section \ref{sec:6}. As it will become clearer there, the explicit form of such a term somehow recalls  the acceleration present inside a spinning mass shell \citep{1984GReGr..16..711M,1985CQGra...2..909P}.

All of them were reobtained from the Kerr metric in harmonic cartesian coordinates as in Equation (15) \citet{2014GReGr..46.1671J}.
\section{The calculational scheme for the orbital effects}\lb{sec:4}
The orbital effects of those accelerations of Section \ref{acc4} which have not yet been explicitly investigated in the literature can be calculated with the standard perturbative techniques in terms of the Keplerian orbital elements. 
They are as follows. The orbit's semimajor axis is\footnote{It should not be confused with the dimensional black hole's spin parameter, usually dubbed  $a_\mathrm{g}$ as in \rfr{a}.} $a$, $\nk:=\sqrt{G M/a^3}$ is the Keplerian mean motion, $P_\mathrm{K} = 2\uppi/\nk$ is the Keplerian orbital period, $e$ is the eccentricity, $p:=a\ton{1-e^2}$ is the semilatus rectum, $I$ is the inclination of the orbital plane to the fundamental plane $\Pi$ of the coordinate system adopted, $\mathit{\Omega}$ is the longitude of the ascending node\footnote{It should not be confused with the angular speed $\Omega$ of a rotating reference frame and with its gravitational analogue $\Omega_\mathrm{g}$ of \rfr{OLT}. }, $\omega$ is the argument of pericentre, $f$ is the true anomaly, $u:=\omega+f$ is the argument of latitude, $\eta$ is the mean anomaly at epoch, and $r$ is the distance of the test particle from the primary which, in the case of the Keplerian ellipse, is 
\eqi
r = \rp{p}{1 + e\cos f}.\lb{rkep}
\eqf

Furthermore, if $\boldsymbol A$ is any additional acceleration with respect of the Newtonian inverse-square one,
it has, first, to be decomposed in three components 
\begin{align}
A_r \lb{Ar}&:= \boldsymbol{A}\boldsymbol\cdot\boldsymbol{\hat{r}},\acap
A_\tau \lb{Atau}& = \boldsymbol{A}\boldsymbol\cdot\boldsymbol{\hat{\tau}},\acap
A_h \lb{Ah}& = \boldsymbol{A}\boldsymbol\cdot\boldsymbol{\hat{h}}
\end{align} 
along the mutually perpendicular radial, transverse and normal directions. The unit vectors of the latter ones are
\begin{align}
\boldsymbol{\hat{r}} \lb{vr}& =\grf{\cos\mathit{\Omega}\cos u - \cos I \sin\mathit{\Omega}\sin u, \sin\mathit{\Omega}\cos u + \cos I \cos\mathit{\Omega}\sin u, \sin I \sin u}, \acap
\boldsymbol{\hat{\tau}} \lb{vt}& =\grf{-\cos\mathit{\Omega}\sin u -\cos I \sin\mathit{\Omega}\cos u, -\sin\mathit{\Omega}\sin u + \cos I \cos\mathit{\Omega}\cos u, \sin I \cos u}, \acap
\boldsymbol{\hat{h}} \lb{vh}& =\grf{\sin I\sin\mathit{\Omega},-\sin I\cos\mathit{\Omega},\cos I}.
\end{align}
The unit vector $\boldsymbol{\hat{h}}$ is directed along the orbital angular momentum perpendicularly to the orbital plane.

Then, \rfrs{Ar}{Ah} must be inserted in the right-hand-sides of the equations for the variations of the orbital elements in the Euler-Gauss form \citep{1961mcm..book.....B,Sof89,2003ASSL..293.....B,1991ercm.book.....B,2005ormo.book.....R,2011rcms.book.....K,2014grav.book.....P,SoffelHan19,2024gpno.book.....I}, which are valid irrespectively of the physical origin of $\bds A$. They are
\begin{align}
\dert{a}{t} \lb{aGa} &= \rp{2}{\nk \sqrt{1-e^2}} \qua{e A_{r} \sin f + \ton{\rp{p}{r}} A_{\tau}},\\ \nonumber \\
\dert e t \lb{eGa} & = \rp{\sqrt{1-e^2}}{\nk a} \grf{A_{r} \sin f + A_{\tau} \qua{\cos f + \rp{1}{e} \ton{1-\rp{r}{a}} }}, \\ \nonumber\\
\dert I t \lb{IGa}& = \rp{1}{\nk a \sqrt{1-e^2}} A_{h} \ton{\rp{r}{a}} \cos u, \\ \nonumber\\
\dert{\mathit{\Omega}} t \lb{OGa}& = \rp{1}{\nk a \sin I \sqrt{1-e^2}} A_{h} \ton{\rp{r}{a}} \sin u, \\ \nonumber\\
\dert \omega t \lb{oGa} & = \rp{\sqrt{1-e^2}}{\nk a e} \qua{-A_{r} \cos f + A_{\tau} \ton{1 + \rp{r}{p}} \sin f} - \cos I \dert{\mathit{\Omega}} t, \\ \nonumber \\
\dert\eta{t} \lb{etaGa} &= -\rp{2}{\nk a} A_{r} \ton{\rp{r}{a}} - \rp{\ton{1-e^2}}{\nk a e} \qua{-A_{r} \cos f +A_{\tau} \ton{1+\rp{r}{p}} \sin f}.
\end{align}

Finally, after having evaluated their right-hand-sides onto the Keplerian ellipse assumed as reference unperturbed trajectory, the average over one orbital period $P_\mathrm{K}$ has to be calculated by means of
\eqi
\rp{dt}{df}=\rp{r^2}{\sqrt{GM p}},
\eqf
where $r$ is given by \rfr{rkep}.

For computational purposes, it is convenient to express the radial and transverse unit vectors of \rfrs{vr}{vt} as
\begin{align}
\bds{\hat{r}} \lb{ur}& = \bds{\hat{l}}\cu +\bds{\hat{m}}\su,\acap
\bds{\hat{\tau}} \lb{ut}& = \bds{\hat{m}}\cu -\bds{\hat{l}}\su.
\end{align}
In them,
\begin{align}
\boldsymbol{\hat{l}} \lb{elle}&:=\grf{\cos\mathit{\Omega},\sin\mathit{\Omega},0},\acap
\boldsymbol{\hat{m}}\lb{emme}&:=\grf{-\cos I\sin\mathit{\Omega},\cos I\cos\mathit{\Omega},\sin I}
\end{align}
are two mutually perpendicular unit vectors lying in the orbital plane such that $\boldsymbol{\hat{l}}$ is directed along the line of nodes\footnote{It is the intersection of the orbital plane with the fundamental plane $\Pi$ of the coordinate system adopted.} towards the ascending node $\ascnode$. The unit vectors $\boldsymbol{\hat{l}},\boldsymbol{\hat{m}},\boldsymbol{\hat{h}}$ form a right-handed orthonormal basis such that $\boldsymbol{\hat{l}}\boldsymbol\times\boldsymbol{\hat{m}} = \boldsymbol{\hat{h}}$ holds, as can be directly verified by using \rfr{vh} and \rfrs{elle}{emme}.

It is useful to define also the following auxiliary quantities specifically pertaining the accelerations of Section \ref{acc4}:
\begin{align}
\mathtt{kl} \lb{kl} & := \boldsymbol{\hat{k}}\boldsymbol\cdot\boldsymbol{\hat{l}},\acap
\mathtt{km} \lb{km} & := \boldsymbol{\hat{k}}\boldsymbol\cdot\boldsymbol{\hat{m}},\acap
\mathtt{kh} \lb{kh} & := \boldsymbol{\hat{k}}\boldsymbol\cdot\boldsymbol{\hat{h}},
\end{align}
and
\begin{align}
\widehat{T}_1 \lb{T1}& := 1,\acap
\widehat{T}_2 \lb{T2}& := \mathtt{kl}^2 + \mathtt{km}^2,\acap
\widehat{T}_3 \lb{T3}& := \mathtt{kl}^2 - \mathtt{km}^2,\acap
\widehat{T}_4 \lb{T4}& := \mathtt{kh}\,\mathtt{kl},\acap
\widehat{T}_5 \lb{T5}& := \mathtt{kh}\,\mathtt{km},\acap
\widehat{T}_6 \lb{T6}& := \mathtt{kl}\,\mathtt{km}.
\end{align}

In spherical coordinates, the components of the primary's spin axis $\bds{\hat{k}}$, assumed aligned with the reference $z$ axis, are
\begin{align}
k_r \lb{kr} & =\cos\theta,\acap
k_\theta \lb{ktheta} &  =-\sin\theta,\acap
k_\phi \lb{kphi} & = 0.
\end{align} 

Useful dimensionless quantities which will be used in the next Sections are 
\eqi
x_\mathrm{s} :=\rp{a}{r_\mathrm{s}},\acap
\eqf
where 
\eqi
r_\mathrm{s}:=\rp{2GM}{c^2}
\eqf
is the hole's Schwarzschild radius, and
\eqi
\mathfrak{m}_\odot := \rp{M}{M_\odot},
\eqf
where $M_\odot$ is the Sun's mass.

A final remark about the interpretation of the outcome of the aforementioned calculational procedure is in order to avoid possible misunderstandings. Indeed, \rfrs{aGa}{etaGa} return the rates of the contact orbital elements some of which, in general, do not coincide with their osculating counterparts when velocity-dependent accelerations are present \citep{2004A&A...415.1187E,2005CeMDA..91...75E,2005NYASA1065..346E,2011rcms.book.....K}. Checking explicitly such a subtle feature in the present case is beyond the scope of the present paper.
\section{The velocity-dependent acceleration of order $\mathcal{O}\ton{S/c^4}$ }\lb{sec:5}
The components of the velocity-dependent acceleration of the order of $\mathcal{O}\ton{S/c^4}$ in spherical coordinates turn out to be
\begin{align}
A_r^{\ton{S/c^4}} \lb{Ar2}& = -\rp{2G S \ton{4 G M + 3 r \dot r^2} \dot\phi \sin^2\theta}{c^4 r^3},\acap
A_\theta^{\ton{S/c^4}} \lb{Atheta2}& = \rp{6G S \dot\phi \sin\theta \ton{2 G M \cos\theta - r^2 \dot r \dot\theta \sin\theta}}{c^4 r^3},\acap
A_\phi^{\ton{S/c^4}} \lb{Aphi2}& = \rp{G S\grf{-12 G M r \dot\theta \cos\theta + \dot r \qua{4 G M - 3 r^3 \dot\phi^2\ton{1-\cos 2\theta}} \sin\theta}}{c^4  r^4}.
\end{align}
\Rfrs{Ar2}{Aphi2} can be cast into the coordinate-independent vector form
\begin{align}
{\bds A}^{\ton{S/c^4}} \lb{Ac}& = \rp{2GS}{c^4}\grf{\rp{2GM}{r^4}\qua{-3\bds{\hat{k}}\bds\times\bds v  + 5\ton{\bds{\hat{k}}\bds\times\bds v\bds\cdot\bds{\hat{r}}}\bds{\hat{r}}  + 4 v_r\bds{\hat{k}}\bds\times\bds{\hat{r}} } + \rp{3}{r^3}\ton{\bds{\hat{k}}\bds\times\bds v\bds\cdot\bds{\hat{r}}}v_r\bds v 
},
\end{align}
valid for an arbitrary orientation of the hole's spin axis $\bds{\hat{k}}$ in space.
%
%
%
The resulting orbital rates of change, averaged over one orbital revolution, are
\begin{align}
\dert a t \lb{dadt2}& = 0,\acap
\dert e t \lb{dedt2}& = 0,\acap
\dert I t \lb{dIdt2}& = -\rp{G^2 S M\grf{e^2\mathtt{km}\sin 2\omega + \mathtt{kl}\qua{3\ton{2+e^2} + e^2\cos 2\omega}}}{c^4 a^4\ton{1-e^2}^{5/2}},\acap
\dert{\mathit{\Omega}} t \lb{dOdt2}& = -\rp{G^2 S M\grf{-\mathtt{km}\qua{-3\ton{2 + e^2} + e^2\cos 2\omega} + e^2\mathtt{kl}\sin 2\omega}}{c^4 a^4\ton{1-e^2}^{5/2}\sin I},\acap
\dert \omega t \lb{dodt2}& = \rp{G^2 S M\ton{2\ton{5 + e^2}\mathtt{kh} +\cot I\grf{-\mathtt{km}\qua{-3\ton{2 + e^2} + e^2\cos 2\omega} + e^2\mathtt{kl}\sin 2\omega}}}{c^4 a^4\ton{1-e^2}^{5/2}},\acap
\dert \eta t \lb{detadt2}& = \rp{6G^2 S M\ton{1 + 2 e^2}\mathtt{kh}}{c^4 a^4\ton{1-e^2}^{2}}.
\end{align}

By using \rfrs{S}{Q2}, valid for a Kerr black hole, the amplitudes of \rfrs{dadt2}{detadt2} turn out to be
\eqi
\dert\qk t^{\ton{S/c^4}}\propto \rp{c^3 \chi}{16\,G M x_\mathrm{s}^4} = \rp{\chi}{\mathfrak{m}_\odot x_\mathrm{s}^4} 2.3\times 10^{13}\,\mathrm{deg\,yr}^{-1},\,\qk= I,\mathit{\Omega},\omega,\eta,\lb{grul}
\eqf
while the shifts per orbit are proportional to
\eqi
\Delta\qk^{\ton{S/c^4}}\propto \rp{\uppi\,\chi}{2\sqrt{2}\,x_\mathrm{s}^{5/2}},\,\qk= I,\mathit{\Omega},\omega,\eta.\lb{grullo}
\eqf

For $e=0$, \rfr{grul} implies that
\begin{align}
\rp{d\qk^{\ton{S/c^4}}/dt}{d\qk^\mathrm{LT}/dt} \lb{b1}& \propto \rp{1}{2x_\mathrm{s}}.
%
%
\end{align}
%
%
%
The Lense-Thirring effect is always larger than its counterpart of  the order of $\mathcal{O}\ton{S/c^4}$, as per \rfr{b1}.
More precisely, if $e=0$, in the case of $I$ and $\mathit{\Omega}$, the figure of \rfr{b1} has to be rescaled by a factor of 3. Indeed, \rfrs{dIdt2}{dOdt2} bring in a scaling factor of 6 for a circular orbit. On the other hand, the corresponding Lense-Thirring amplitudes for the same value of the eccentricity are rescaled by a factor of 2, as per Equations (5.32)-(5.33) of \citet{2024gpno.book.....I}, being the factors accounting for the spin-orbit configuration equal  for both the spin-dependent effects considered.
%
Thus, if $x_\mathrm{s}\simeq 7$ \citep{2025PhRvD.111d4035I}, as in the case of the disk/jet precession around M87$^\ast$ \citep{2023Natur.621..711C}, the precessional contribution of the order of $\mathcal{O}\ton{S/c^4}$ would amount to $\simeq 21\%$ of the Lense-Thirring effect itself. Since the current accuracy level in measuring the precessional rate in M87$^\ast$, attributed so far to the Lense-Thirring effect, is $\simeq 3\%$ \citep{2023Natur.621..711C}, it does not seem unrealistic to think that such a novel feature of motion will become detectable in the not too distant future. Essentially the same situation occurs for  the recently measured disk/jet precession associated with the tidal disruption event AT2020afhd \citep{TDELTCinesi2025}, also attributed to the Lense-Thirring effect. Indeed, while its measurement accuracy is of the order of $7\%$ \citep{TDELTCinesi2025}, on the other hand, it can be shown that the dimensionless radius of the fictitious test particle orbit effectively modeling it is about $x_\mathrm{s}\simeq 8$, implying that the effect of the order of $\mathcal{O}\ton{S/c^4}$ amounts to $\simeq 18\%$ of the Lense-Thirring precession.
\section{The position-dependent acceleration of the order of $\mathcal{O}\ton{S^2/c^4}$}\lb{sec:6}
The position-dependent part of the acceleration of the order of $\mathcal{O}\ton{S^2/c^4}$ which does not have a counterpart of the order of $\mathcal{O}\ton{J_2/c^2}$ in the case of an oblate material primary may be dubbed as Kerr-only quadrupolar acceleration, or Kerr-only $Q_2/c^2$ acceleration because of \rfr{Q2}.  
It can be formally expressed as the sum of two terms apparently resembling a centrifugal and a centripetal acceleration, respectively, as
\eqi
{\bds{A}}^{\ton{S^2/c^2}} = -{\bds\Omega}_\mathrm{g}\bds\times{\bds\Omega}_\mathrm{g}\bds\times\bds r - \Omega_\mathrm{g}^2\bds r = -\ton{{\bds\Omega}_\mathrm{g}\bds\cdot\bds r}{\bds\Omega}_\mathrm{g}, \lb{Accel}
\eqf
where ${\bds\Omega}_\mathrm{g}$ is given by 
\eqi
{\bds\Omega}_\mathrm{g} = \rp{GS}{c^2 r^3}\qua{\bds{\hat{k}} -3\ton{\bds{\hat{k}}\bds\cdot\bds{\hat{r}}}\bds{\hat{r}}}.\lb{OLT}
\eqf
It is intended that \rfr{Accel} concurs with \rfr{A1pNJ2} to form the total acceleration of the order of $\mathcal{O}\ton{S^2/c^4}$ of \rfrs{Ar1}{Aphi1}, which generally depends on the velocity as well. It may be interesting to note that \rfr{Accel} somehow recalls the acceleration experienced by a test particle in the interior of a rotating mass shell \citep{1984GReGr..16..711M,1985CQGra...2..909P}.

The explicit expressions of the components of \rfr{Accel} in spherical coordinates  
\begin{align}
A^{\ton{S^2/c^2}}_r & = -\rp{4G^2 S^2\cos^2\theta}{c^4 r^5},\acap
A^{\ton{S^2/c^2}}_\theta & = \rp{G^2 S^2\sin 2\theta}{c^4 r^5},\acap
A^{\ton{S^2/c^2}}_\phi & = 0
\end{align}
elucidate that, in fact, the attributes \virg{centrifugal} and \virg{centripetal} are merely formal. Indeed, it can be noted that  it vanishes in the hole's equatorial plane for $\theta=\qp/2$, while it is directed inward along the rotation axis for $\theta=0,\qp$.
%
%
%
%
%
%

The resulting orbital rates of change, averaged over one orbital revolution, are
\begin{align}
\dert a t \lb{dadt}& = 0,\acap
\dert e t \lb{dedt} & = -\rp{e G^2 S^2\ton{\widehat{T}_3\sin 2\omega - 2\widehat{T}_6\cos 2\omega}}{4c^4 \nk a^6 \ton{1-e^2}^2},\acap
\dert I t \lb{dIdt} & = \rp{G^2 S^2\grf{\widehat{T}_4\qua{2\ton{2+e^2} + e^2\cos 2\omega} + e^2\widehat{T}_5\sin 2\omega}}{4c^4 \nk a^6 \ton{1-e^2}^3},\acap
\dert{\mathit{\Omega}} t \lb{dOdt} & = \rp{G^2 S^2\grf{e^2\widehat{T}_4\sin 2\omega + \widehat{T}_5\qua{2\ton{2+e^2} - e^2\cos 2\omega}}}{4c^4\nk a^6 \sin I \ton{1-e^2}^3},\acap
\dert \omega t \nonumber & = \rp{G^2 S^2}{8c^4\nk a^6 \ton{1-e^2}^3}\grf{
6 \ton{4 + e^2}\widehat{T}_2 + \ton{2 + 3 e^2}\widehat{T}_3\cos 2\omega - 2 e^2\cot I\widehat{T}_4\sin 2\omega +\right.\acap
\lb{dodt} & + \left. \cot I\widehat{T}_5\qua{-4\ton{2 + e^2} +2e^2\cos 2\omega } + 2\ton{2 + 3 e^2}\widehat{T}_6\sin 2\omega
},\acap
\dert \eta t \lb{detadt} & = \rp{G^2 S^2\qua{2 \ton{4 + 5 e^2}\widehat{T}_1 + \ton{-2 + 5 e^2}\ton{\widehat{T}_3\cos 2\omega + 2\widehat{T}_6\sin 2\omega}}  }{8c^4\nk a^6 \ton{1-e^2}^{5/2}}.
\end{align}
By using \rfrs{S}{Q2}, valid for a Kerr black hole, the amplitudes of Equations (8.9)-(8.10) in \citet{2024gpno.book.....I} and of \rfrs{dadt}{detadt} turn out to be
\begin{align}
\dert{\qk}{t}^{\ton{S^2/c^2}} \lb{gross2}& \propto \rp{c^3\chi^2}{16\sqrt{2}\,GM\,x_\mathrm{s}^{9/2}} = \rp{\chi^2}{\mathfrak{m}_\odot\,x_\mathrm{s}^{9/2}}1.6\times 10^{13}\,\mathrm{deg}\,\mathrm{yr}^{-1},\,\qk = a,e,I,\mathit{\Omega},\omega,\eta,
\end{align}
while the shifts per orbit are proportional to
\begin{align}
\Delta\qk^{\ton{S^2/c^2}} &\propto \rp{\uppi\,\chi^2}{4\,x_\mathrm{s}^{3}}.
\end{align}
For $e=0$, \rfr{gross2} implies that
\eqi
\rp{d\qk^{\ton{S^2/c^4}}/dt}{d\qk^\mathrm{LT}/dt} = \rp{\chi}{2\sqrt{2}x_\mathrm{s}^{3/2}}\lb{w1};
\eqf
the Lense-Thirring precessions are even larger than their counterparts of the order of $\mathcal{O}\ton{S^2/c^4}$. In the case of $I$ and $\mathit{\Omega}$, the figure of \rfr{w1} has to be rescaled by a factor of $\ton{11/4}\mathtt{kh}$. Indeed, Equations (8.9)-(8.10) of \citet{2024gpno.book.....I} and \rfrs{dIdt}{dOdt} bring in an overall scaling factor of $9/2+1 = 11/2$ for a circular orbit. On the other hand, the corresponding Lense-Thirring amplitudes for the same value of the eccentricity are rescaled by a factor of 2, as per Equations (5.32)-(5.33) of \citet{2024gpno.book.....I}. Thus, if $x_\mathrm{s}\simeq 7$ \citep{2025PhRvD.111d4035I}, as in the case of the disk/jet precession around M87$^\ast$ \citep{2023Natur.621..711C}, the total precessional contribution of the order of $\mathcal{O}\ton{S^2/c^4}$ would amount to $\lesssim 5\%$ of the Lense-Thirring effect itself. Since the current accuracy level in measuring the precessional rate in M87$^\ast$, attributed so far to the Lense-Thirring effect, is $\simeq 3\%$ \citep{2023Natur.621..711C}, it does not appear impossible  that such a feature of motion can be  already detectable. Maybe, they will become even better measurable in the not too distant future in view of possible improvements in the measuring accuracy of the disk/jet precession.
\section{The velocity-dependent acceleration of order $\mathcal{O}\ton{S^3/c^4}$ }\lb{sec:7}
The components of the velocity-dependent acceleration of the order of $\mathcal{O}\ton{S^3/c^4}$ in spherical coordinates turn out to be
\begin{align}
A_r^{\ton{S^3/c^4}} \lb{Ar3}& = -\rp{G S^3 \dot\phi \ton{3 + 5 \cos 2\theta} \sin^2\theta}{2c^4 M^2 r^4},\acap
A_\theta^{\ton{S^3/c^4}} \lb{Atheta3}& = \rp{G S^3 \dot\phi \ton{-2 \sin 2\theta + 5 \sin 4\theta}}{2c^4 M^2 r^4},\acap
A_\phi^{\ton{S^3/c^4}} \lb{Aphi3}& = \rp{G S^3\qua{-4 r \dot\theta \ton{3 \cos\theta + 5 \cos 3\theta} + 3 \dot r \ton{\sin\theta + 5 \sin 3\theta}}}{4c^4 M^2 r^5}.
\end{align}
\Rfrs{Ar3}{Aphi3} can be cast into the coordinate-independent vector form
\begin{align}
{\bds A}^{\ton{S^3/c^4}} \lb{Acc}& = \rp{3GS^3}{c^4 M^2 r^5}\qua{
\ton{-1 + 5 r_k^2} \bds{\hat{k}}\bds\times\bds v + 5 r_k\ton{1 - \rp{7}{3} r_k^2} \bds{\hat{r}}\bds\times\bds v
},
\end{align}
valid for an arbitrary orientation of the hole's spin axis $\bds{\hat{k}}$ in space.
The resulting orbital rates of change, averaged over one orbital revolution, are
\begin{align}
\dert a t \lb{dadt3} & = 0,\acap
\dert e t \lb{dedt3} & = \rp{15 e GS^3\mathtt{kh}\ton{\widehat{T}_3 \sin 2\omega -2 \widehat{T}_6 \cos 2\omega}}{4 c^4 M^2 a^5\ton{1-e^2}^{5/2}},\acap
\dert I t \nonumber & = -\rp{3GS^3}{8 c^4 M^2 a^5\ton{1-e^2}^{7/2}}\grf{2 \ton{2 + 3 e^2}\mathtt{kl}\ton{-4 + 5\widehat{T}_2} + \right.\acap
\lb{dIdt3} &\left. + 5 e^2 \qua{\mathtt{kl} \ton{-2 + 3 \mathtt{kl}^2 + \mathtt{km}^2} \cos 2\omega + 2\mathtt{km}\ton{-1 + 2 \mathtt{kl}^2 + \mathtt{km}^2} \sin 2\omega}},\acap
\dert{\mathit{\Omega}}t \nonumber & = -\rp{3GS^3}{8 c^4 M^2 a^5\ton{1-e^2}^{7/2}\sin I}\grf{2 \ton{2 + 3 e^2}\mathtt{km}\ton{-4 + 5\widehat{T}_2} + \right.\acap
\lb{dOdt3} &\left. + 5 e^2 \qua{-\mathtt{km} \ton{-2 + \mathtt{kl}^2 + 3 \mathtt{km}^2} \cos 2\omega + 2\mathtt{kl}\ton{-1 + \mathtt{kl}^2 + 2 \mathtt{km}^2} \sin 2\omega}},\acap
\dert{\omega} t  \nonumber & = \rp{3GS^3}{8 c^4 M^2 a^5\ton{1-e^2}^{7/2}}\grf{
4 \ton{3 + 2 e^2}\mathtt{kh}\ton{-2 + 5\widehat{T}_2} + 2 \ton{2 + 3 e^2}\mathtt{km}\ton{-4 + 5\widehat{T}_2} \cot I + \right.\acap
\nonumber &\left. + 5 \cos 2\omega \qua{2 \ton{1 + 2 e^2}\mathtt{kh}\,\widehat{T}_3 - e^2\mathtt{km} \ton{-2 + \mathtt{kl}^2 + 3 \mathtt{km}^2} \cot I} + \right.\acap
\lb{dodt3} &\left. + 10\,\mathtt{kl}\qua{2 \ton{1 + 2 e^2} \widehat{T}_5 + e^2 \ton{-1 + \mathtt{kl}^2 + 2 \mathtt{km}^2} \cot I} \sin 2\omega
},\acap
\dert \eta t \lb{detadt3} & = -\rp{3GS^3\mathtt{kh}\qua{5 \widehat{T}_3 \cos 2\omega + 2 \ton{-2 + 5\widehat{T}_2 + 5 \widehat{T}_6 \sin 2\omega}}}{4 c^4 M^2 a^5\ton{1-e^2}^2}.
\end{align}

By using \rfrs{S}{Q2}, valid for a Kerr black hole, the amplitudes of \rfrs{dadt3}{detadt3} turn out to be
\eqi
\dert\qk t^{\ton{S^3/c^4}}\propto \rp{c^3 \chi^3}{32\,G M x_\mathrm{s}^5} = \rp{\chi^3}{\mathfrak{m}_\odot x_\mathrm{s}^5} 1.1\times 10^{13}\,\mathrm{deg\,yr}^{-1},\,\qk= I,\mathit{\Omega},\omega,\eta,\lb{grul2}
\eqf
while the shifts per orbit are proportional to
\eqi
\Delta\qk^{\ton{S^3/c^4}}\propto \rp{\uppi\,\chi^3}{4\sqrt{2}\,x_\mathrm{s}^{7/2}},\,\qk = I,\mathit{\Omega},\omega,\eta.
\eqf

For $e=0$, \rfr{grul2} implies that
\begin{align}
\rp{d\qk^{\ton{S^3/c^4}}/dt}{d\qk^\mathrm{LT}/dt} \lb{u1}& \propto \rp{\chi^2}{4 x_\mathrm{s}^2}.
%
%
%
%
\end{align}
From \rfr{u1} it turns out that the precessions of the order of $\mathcal{O}\ton{S^3/c^4}$ are quite smaller than the Lense-Thirring ones. Thus, it is difficult but not impossible that such features of motion  can play a role  in the interpretation of the observations of disk/jet precessions around supermassive black holes.
\section{The impact on the M87$^\ast$ disk/jet precession}
\begin{figure}[ht!]
\centering
\begin{tabular}{cc}
\includegraphics[width = 7.5 cm]{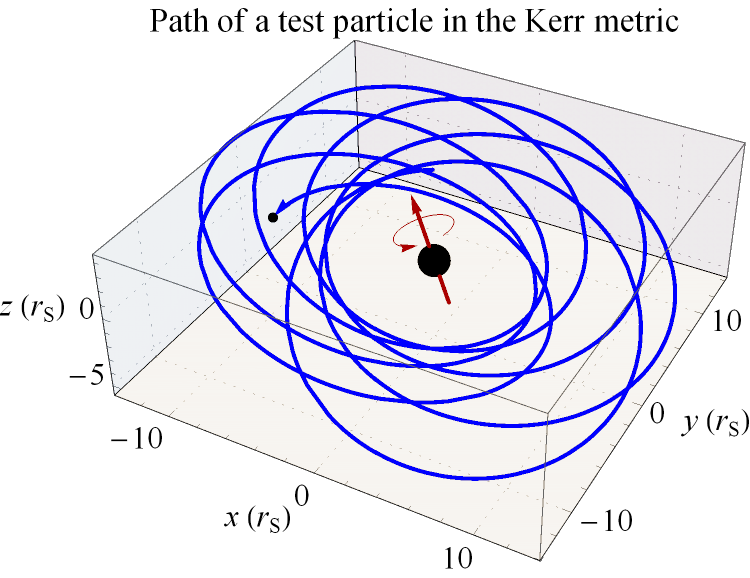} & \includegraphics[width = 7.5 cm]{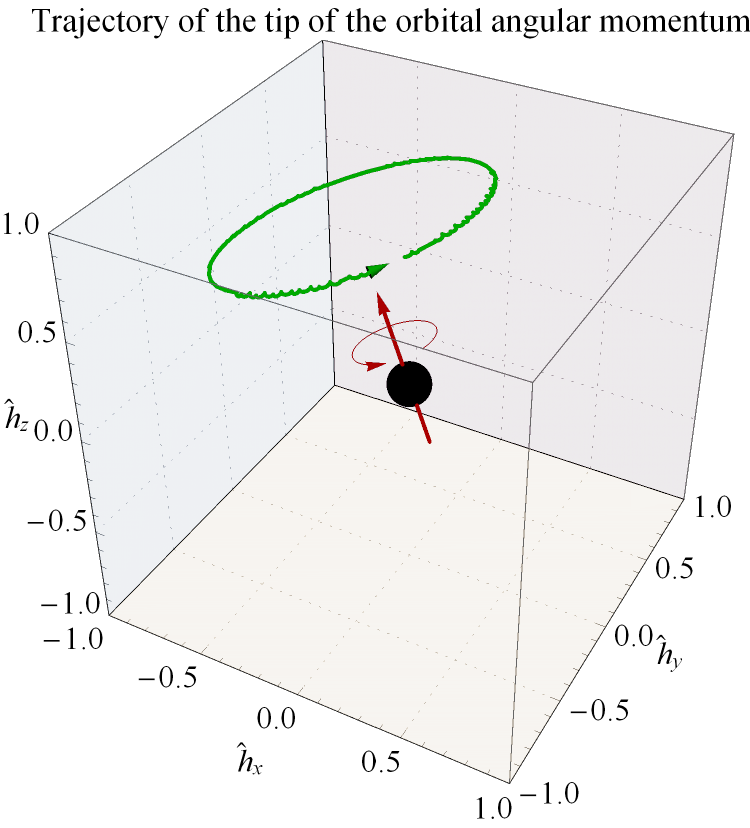}\\
\end{tabular}
\caption{Numerically integrated trajectories of the motion of a fictional test particle (blue) and of the tip of its orbital angular momentum (green) in the Kerr metric of a putative rotating supermassive black hole with $M = 6.5\times 10^9 M_\odot,\,\chi=0.98$; the harmonic coordinates are used. The hole's spin axis  (dark red) is parameterized as ${\hat{k}}_x=\cos\delta\cos\eta,\,{\hat{k}}_y=\cos\delta\sin\eta,\,{\hat{k}}_x=\sin\delta$   with $\delta = 70^\circ,\,\eta = 195^\circ$. The initial values of the particle's Keplerian orbital elements are $a_0=6.45\,r_\mathrm{S},\,e_0=0.0,\,I_0 = 25^\circ,\,\mathit{\Omega}_0=26^\circ,\,\omega_0 = 43^\circ,\,f_0 = 50^\circ$, so that the angle $\psi$ between $\bds{\hat{k}}$ and the particle's orbital angular momentum is $\psi = 34.5^\circ$. All the formally 1pN and 2pN mass and spin-dependent accelerations arising from the Kerr metric are included. The integrations for the trajectories of the particle's motion and of its orbital angular momentum cover 20 orbital periods and $28.4$ yr, respectively.}\label{Fig:1}
\end{figure}

\lb{sec:8}
The disk/jet coprecession around M87$^\ast$ was recently detected to a $\simeq 3\%$ accuracy \citep{2023Natur.621..711C}. In particular, its measured precessional angular speed $\Omega_\mathrm{p}$ turned out to be \citep{2023Natur.621..711C}
\eqi
\Omega_\mathrm{p}^\mathrm{meas} = 0.56\pm 0.02\,\mathrm{rad\,yr}^{-1}=  32\pm 1\,\mathrm{deg\, yr}^{-1}.\lb{Op}
\eqf
It was recently shown that \rfr{Op} can be reproduced by the Lense-Thirring precessional angular speed of the orbital angular momentum of a fictitious test particle moving along an effective inclined and circular orbit at about 14 gravitational radii from the central black hole \citep{2025PhRvD.111d4035I}. 
The precession of \rfr{vh} is directly related to the rates of change of the inclination $I$ and the node $\mathit{\Omega}$ through
\eqi
\dert{\bds{\hat{h}}}{t} =\derp{\bds{\hat{h}}}{I}\dert I t + \derp{\bds{\hat{h}}}{{\mathit{\Omega}}}\dert{\mathit{\Omega}}t = {\bds{\Omega}}_\mathrm{p}\bds\times\bds{\hat{h}}.\lb{rateh}
\eqf 
In \rfr{rateh}, the theoretical angular velocity ${\bds{\Omega}}_\mathrm{p}$ can be calculated from the expressions of $dI/dt$ and $d\mathit{\Omega}/dt$ for each of the accelerations considered in the previous Sections. 
Figure~\ref{Fig:1} illustrates this approach by depicting the numerically integrated trajectories of a fictional test particle and of the tip of its orbital angular momentum around a hypothetical spinning supermassive black hole with the same mass and spin of, say, M87$^\ast$ by including all the formally 1pN and 2pN accelerations considered so far in the orbiter's dynamics. Both the hole's spin and the orbital configurations are chosen arbitrarily.

\begin{figure}[ht!]
\centering
\begin{tabular}{cc}
\includegraphics[width = 7.5 cm]{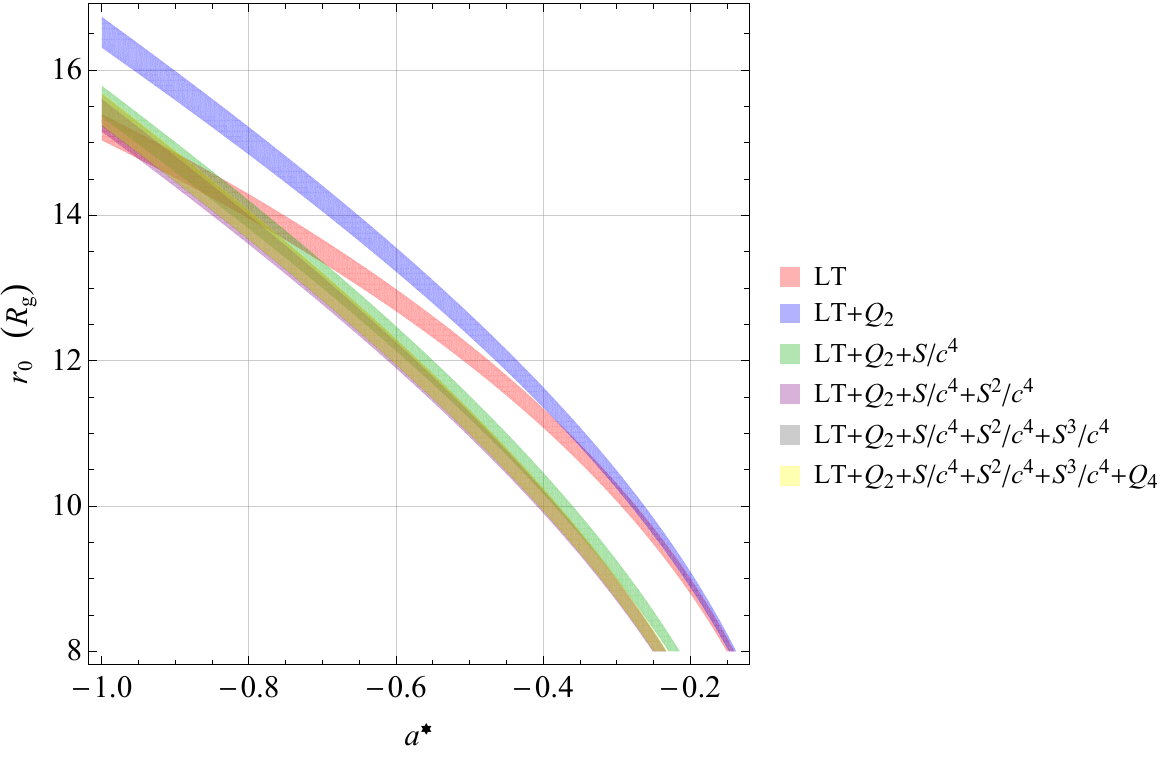} & \includegraphics[width = 7.5 cm]{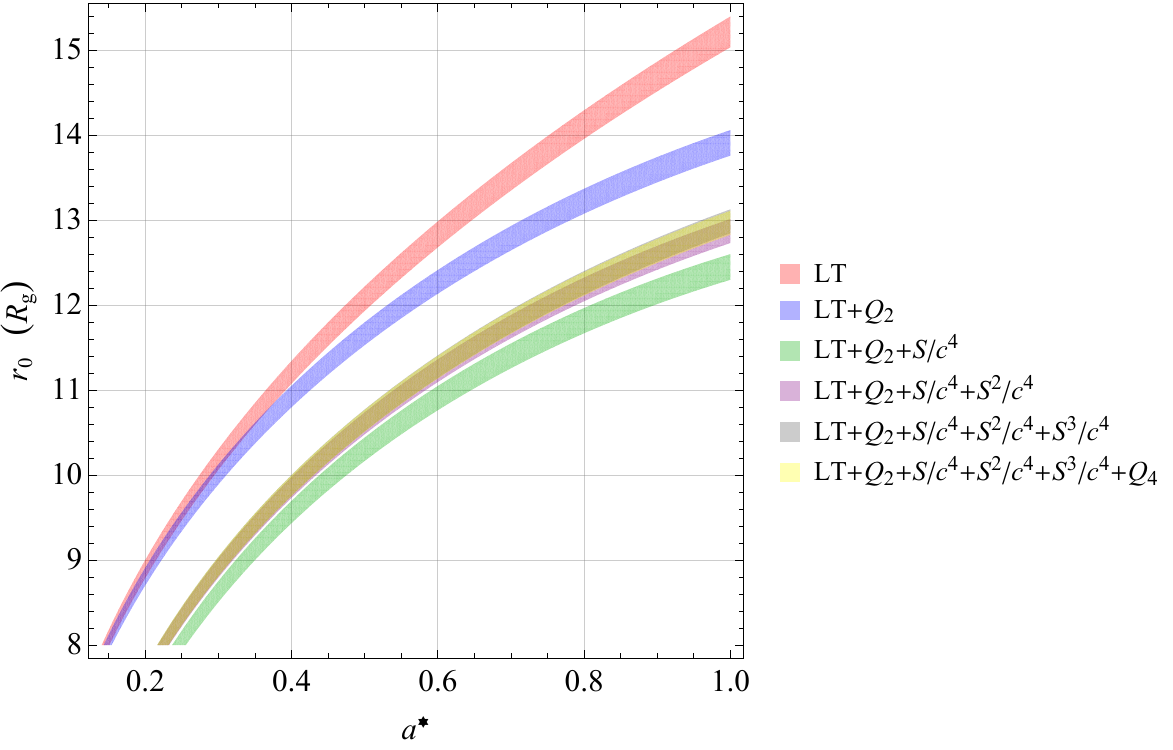}\\
\includegraphics[width = 7.5 cm]{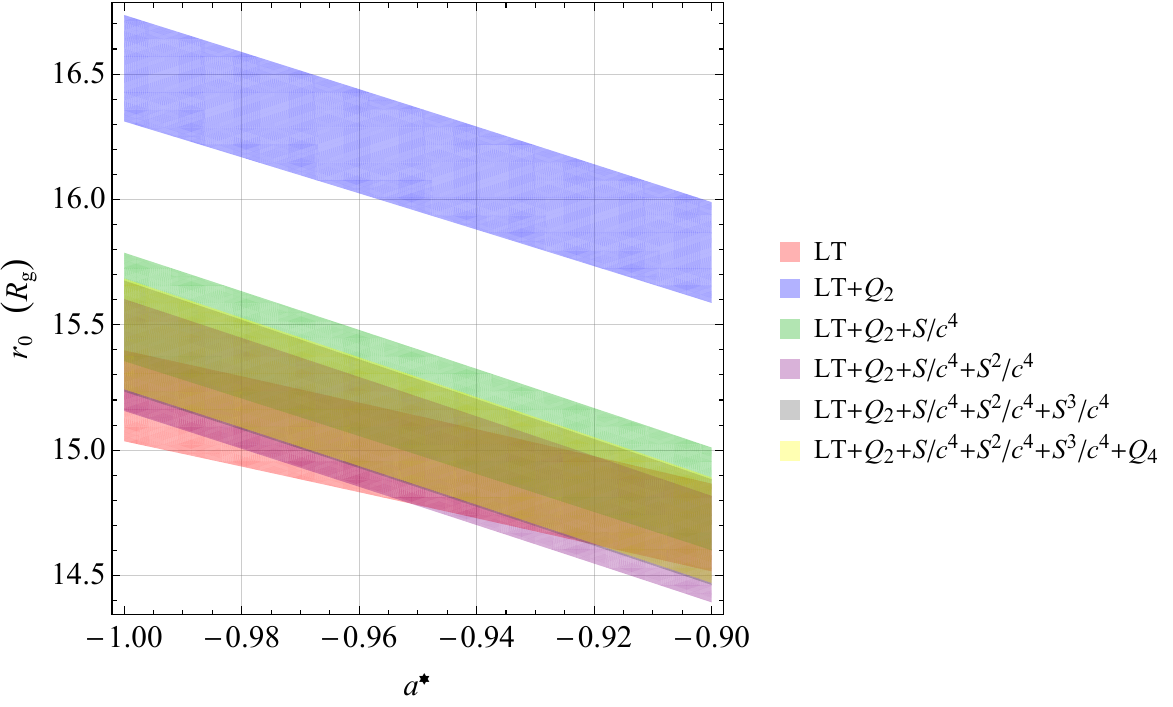} & \includegraphics[width = 7.5 cm]{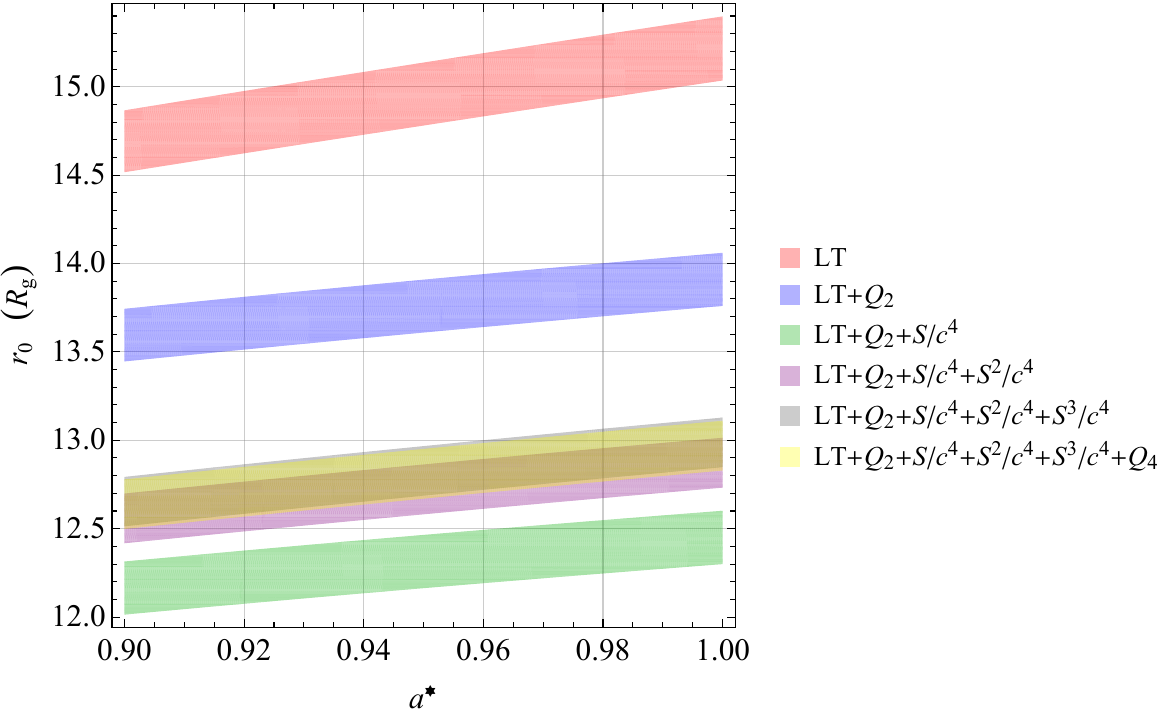}\\

\end{tabular}
\caption{Allowed regions in the $\grf{a^\ast,r_0}$ parameter space of M$87^\ast$ obtained by imposing that the absolute values of the theoretical  frequencies of the precession of the orbital plane of a fictitious test particle moving along a tilted, circular effective orbit around the supermassive black hole lie within the experimental range of \rfr{Op}. The labels \virg{LT}, \virg{$Q_2$} and \virg{$Q_4$} denote the standard Lense-Thirring and the formally Newtonian even zonal harmonic frequencies of degree $\ell=2,4$, respectively, while \virg{$S/c^4$}, \virg{$S^2/c^4$} and \virg{$S^3/c^4$}  refer to the contributions proportional to $S/c^4$, $S^2/c^4$ (the total one) and $S^3/c^4$, respectively. See Figure~\ref{Fig:3} for an overview of all permitted regions.}\label{Fig:2}
\end{figure}

\begin{figure}[ht!]
\centering
\begin{tabular}{c}
\includegraphics[width = 15 cm]{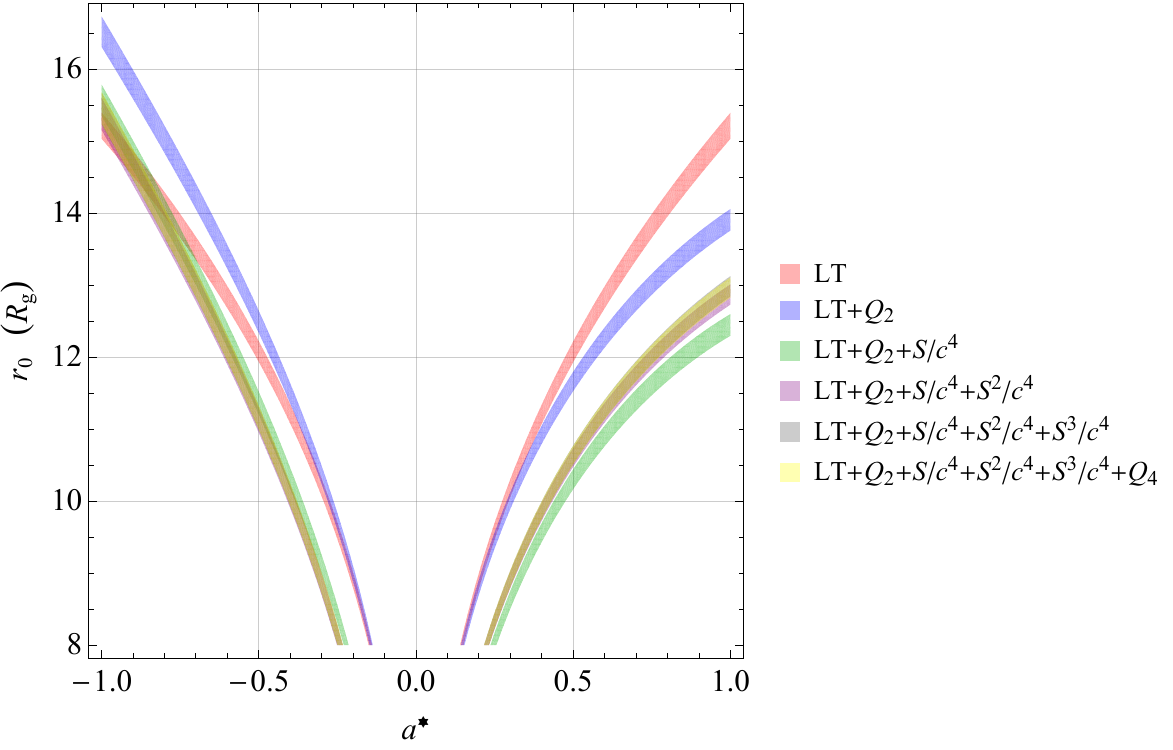} \\

\end{tabular}
\caption{Joint representation of all the allowed regions in the $\grf{a^\ast,r_0}$ parameter space of M$87^\ast$ obtained by imposing that the absolute values of the theoretical  frequencies of the precession of the orbital plane of a fictitious test particle moving along a tilted, circular effective orbit around the supermassive black hole lie within the experimental range of \rfr{Op}. The labels \virg{LT}, \virg{$Q_2$} and \virg{$Q_4$} denote the standard Lense-Thirring and the formally Newtonian even zonal harmonic frequencies of degree $\ell=2,4$, respectively, while \virg{$S/c^4$}, \virg{$S^2/c^4$} and \virg{$S^3/c^4$}  refer to the contributions proportional to $S/c^4$, $S^2/c^4$ (the total one) and $S^3/c^4$, respectively.}\label{Fig:3}
\end{figure}

Figures~\ref{Fig:2} to \ref{Fig:3} display the allowed regions in the two-dimensional parameter space spanned by the dimensionless spin parameter of M87$^\ast$, here dubbed $a^\ast$, and the effective radius $r_0$, in units of gravitational radii, of the orbit of the fictitious test particle. 

They were obtained as in \citet{2025PhRvD.111d4035I}, namely by imposing that the absolute value of the analytically computed theoretical precessional frequencies, considered as functions of $a^\ast$ and $r_0$, is constrained by the experimental range of \rfr{Op}.

It can be noticed that, for positive values of $a^\ast$, the inclusion of the contributions of the order of $\mathcal{O}\ton{S/c^4}$ and, to a lesser extent, of $\mathcal{O}\ton{S^2/c^4}$ as well, change the allowed regions in a non-negligible way, leading generally to smaller permitted values of $r_0$. The further addition of the term of the order of $\mathcal{O}\ton{S^3/c^4}$ has an almost negligible impact, while the formally Newtonian term proportional to $Q_4$ has virtually no effect at all. The negative values of $a^\ast$ yield to a generally different pattern of the allowed regions implying larger values of $r_0$. Furthermore, for $\left|a^\ast\right|\simeq 0.9-1$, all the permitted domains tend to overlap apart from the one obtained by including only the Lense-Thirring and the quadrupolar terms.

\begin{figure}[ht!]
\centering
\begin{tabular}{cc}
\includegraphics[width = 7.5 cm]{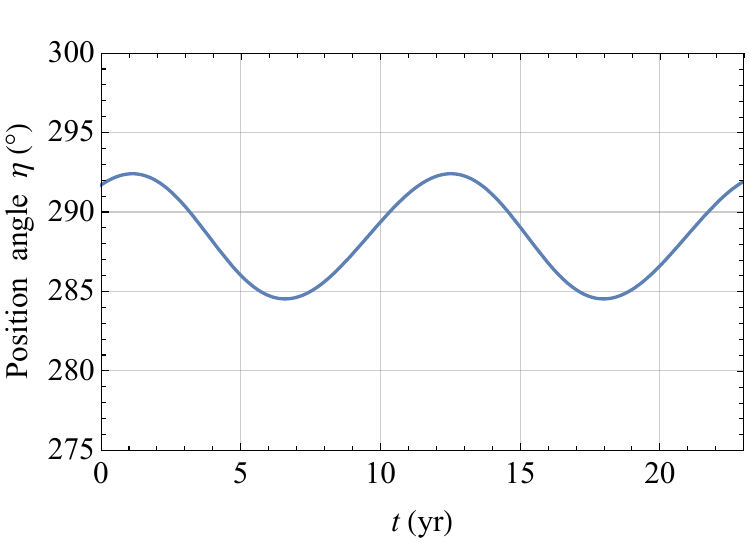} \\
\includegraphics[width = 7.5 cm]{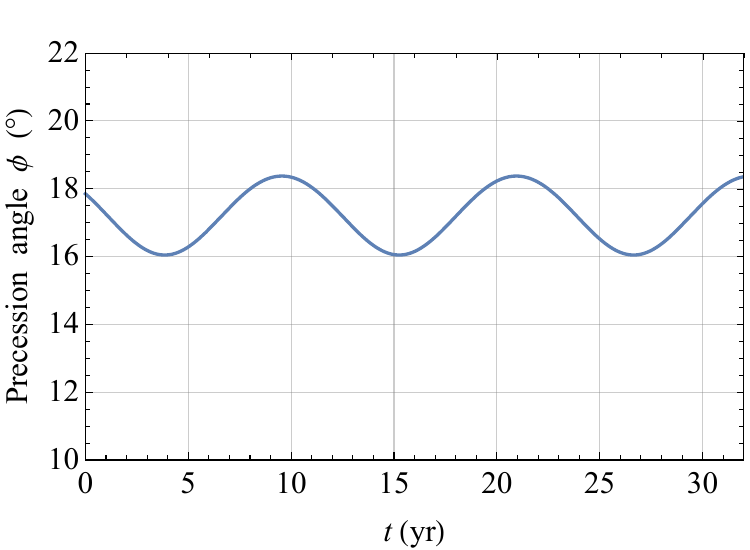} \\

\end{tabular}
\caption{Numerically produced time series, in
degrees, for the position angle $\eta$ (upper panel) and the precession angle $\phi$ (lower panel), in the notation of \citet{2023Natur.621..711C}. The accelerations included in the numerical integrations are the Lense-Thirring and the formally Newtonian quadrupolar ones along with the 2pN spin-dependent accelerations proportional to $S/c^4, S^2/c^4, S^3/c^4$. The values $a=0.98,
r_0 = 13.1 R_\mathrm{g}$, and $M = 6.5\times 10^9 M_\odot$ were used for
the simultaneous numerical integration along
with $\phi_0 = 17.85^\circ$ and $\eta_0 = 291.7^\circ$, retrieved from Fig. 2(b) and
extended data Fig. 4 of \citet{2023Natur.621..711C}, for the initial conditions of $\phi$ and $\eta$.
The time spans and the ranges of values on the vertical axes of
both panels have the same length as those in Fig. 2 and extended
data Fig. 4 of \citet{2023Natur.621..711C} for a better comparison with the latter ones.}\label{Fig:4}
\end{figure}

As a further confirmation of the validity of the approach followed here, Figure~\ref{Fig:4} displays the numerically produced time series of the orbital angular momentum's position and precession angles $\eta$ and $\phi$, respectively, in the notation of \citet{2023Natur.621..711C}. The dynamical models adopted included all the accelerations investigated so far and the values $a_\ast=0.98,\,r_0=13.1 R_\mathrm{g}$ retrieved from the allowed regions of Figure~\ref{Fig:2} for the aforementioned dynamics.
It can be noted that the resulting signals shows an excellent agreement with the correspondent experimental curves in \citet{2023Natur.621..711C}.
\section{Summary and overview}\lb{sec:9}
The Kerr metric describing the exterior spacetime of a spinning black hole was written by explicitly expressing the Boyer-Lindquist coordinates $x^0_\mathrm{BL},r_\mathrm{BL},\theta_\mathrm{BL},\phi_\mathrm{BL}$ as functions of the  harmonic ones $x^0,r,\theta,\phi$. Exact expressions for the spacetime metric tensor in terms of the latter ones were, thus, obtained. By expanding $g_{\qm\qn}$ in powers of the two characteristic lengths $\mg$ and $\ag$ of the Kerr metric proportional to the hole's mass $M$ and angular momentum $S$, respectively, approximated formulas for the metric coefficients were derived and used to calculate the geodesic equations of motion of an uncharged,  spinless body orbiting  the hole at some gravitational radii. Accelerations formally up to the 2pN order were calculated and displayed in a coordinate-independent, vector form valid for any orientations of the hole's spin axis $\bds{\hat{k}}$ in space and suitable for practical applications. 

Those of the order of $\mathcal{O}\ton{1/c^2}$ are the position-dependent, formally Newtonian acceleration induced by the mass quadrupole moment  of the primary, and the  velocity-dependent Schwarzschild-like, mass-only and Lense-Thirring accelerations, respectively. They are identical to their counterparts arising in the field of an rotating material body endowed with axial symmetry. 

To the order of $\mathcal{O}\ton{1/c^4}$, there are five accelerations. One of them is the known velocity-dependent 2pN mass-only term. The acceleration of the order of $\mathcal{O}\ton{S^2/c^2}$ can be formally split as the sum of three components. One of them corresponds to the velocity-dependent 1pN quadrupolar acceleration of the order of $\mathcal{O}\ton{J_2/c^2}$ caused by an oblate material body endowed with dimensionless quadrupole mass moment $J_2$. Instead, the other two position-dependent components formally resemble to a centrifugal and a centripetal acceleration in an accelerated rotating frame whose angular velocity vector ${\bds\Omega}_\mathrm{g}$ is, however, a non-uniform, position-dependent vector field. The position-dependent acceleration proportional to $S^4/c^4$ formally corresponds to the Newtonian one due to the second even zonal harmonic multipole $J_4$ of an axisymmetric material primary.
Finally, apparently without a counterpart in the case of a material body, one finds two velocity-dependent accelerations proportional to $S/c^4$ and $S^3/c^4$, respectively. 

The long-term rates of change of the Keplerian orbital elements induced by the accelerations of the order of $\mathcal{O}\ton{S/c^4}$,   $\mathcal{O}\ton{S^3/c^4}$ and by the one of the order of $\mathcal{O}\ton{S^2/c^4}$ sourced by ${\bds\Omega}_\mathrm{g}$ were analytically calculated with standard perturbative techniques. The resulting expressions, averaged over one orbital revolution, are not restricted to any values of the particle's eccentricity $e$ and inclination $I$, and the hole's spin axis $\bds{\hat{k}}$ retains a general orientation in space.

It turned out that the effects of the order of $\mathcal{O}\ton{S/c^4}$ may be measurable by exploiting  the disk/jet coprecessions around supermassive black holes in galactic nuclei, recently measured to a few per cent accuracy as in the case of M87$^\ast$ and the tidal disruption event AT2020afhd. Indeed, it was demonstrated that such a phenomenology can be explained in terms of the Lense-Thirring precession of the orbital angular momentum of a fictitious test particle moving at about fifteen gravitational radii from the central black hole. To this aim, the precession induced by the aforementioned acceleration proportional to $S/c^4$ amounts to about twenty per cent of the measured one, assumed due to the Lense-Thirring effect. Also the total precession of the order of $\mathcal{O}\ton{S^2/c^4}$ is measurable, although to a lesser extent. Instead, the contribution of the order of $\mathcal{O}\ton{S^3/c^4}$ is marginally detectable, if any, while the formally Newtonian one induced by the second even zonal harmonic of the primary is negligible. Such conclusions were graphically confirmed by the displayed allowed domains in the two-dimensional parameter space of the jet precession around M87$^\ast$ spanned by the spin parameter of the latter and the effective orbital radius of the fictitious test particle representing the accretion disk.  
\appendix
\numberwithin{equation}{section}

\appendixsection{}\lb{app:A}
Here, the task outlined in Section \ref{sec:2}, namely the transformation of the Kerr metric from the Boyer-Lindquist coordinates to the harmonic ones, is implemented.

This is done starting from \citep{1997PhRvD..56.4775C,2008PhRvD..77j4001H}
\begin{align}
x^0 \lb{eq0}&=x_\mathrm{BL}^0,\acap
x + i y \lb{eq1}&=\ton{r_\mathrm{BL} - \mg + i a_\mathrm{g}} e^{i\xi_\mathrm{BL}} \sin\theta_\mathrm{BL},\acap
r^2 \lb{eq2}& = \ton{r_\mathrm{BL} - \mg}^2 + a_\mathrm{g}^2\sin^2\theta_\mathrm{BL}, \acap
r^2\cos^2\theta \lb{eq3}& =  \ton{r_\mathrm{BL} - \mg}^2\cos^2\theta_\mathrm{BL}.
\end{align}
In \rfr{eq1}, 
\begin{align}
x \lb{xH}& = r\sin\theta\cos\phi,\acap
y \lb{yH}& =r\sin\theta\sin\phi
\end{align}
are the harmonic Cartesian coordinates, 
\eqi
i:=\sqrt{-1}
\eqf 
is the imaginary unit, and
\eqi
\xi_\mathrm{BL} := \phi_\mathrm{BL} + \rp{a_\mathrm{g}}{r_{+} - r_{-}}\ln\left|\rp{r_\mathrm{BL} - r_{+}}{r_\mathrm{BL} - r_{-}}\right|,\lb{xi}
\eqf
where
\eqi
r_{\pm}:= \mg\pm\sqrt{\mg^2-a_\mathrm{g}^2}.\lb{rpm}
\eqf
In the Kerr metric, $r_\pm$ range within
\begin{align}
\mg &\leq r_{+}\leq 2\mg, \acap
0&\leq r_{-}\leq \mg.
\end{align}
Since in the scenario of interest here the test particle stays far from the event horizon, $r_\mathrm{BL}> r_{\pm}$, and the absolute value in \rfr{xi} can be dropped in the following.

From \rfrs{eq2}{eq3}, two solutions for $\sin^2\theta_\mathrm{BL}$ can be obtained in terms of the harmonic coordinates $r$ and $\theta$. They are
\eqi
\sin^2\theta_\mathrm{BL} = \rp{a_\mathrm{g}^2 + r^2 \mp\sqrt{\mathcal{B}_1}}{2a_\mathrm{g}^2},
\eqf
where
\eqi
\mathcal{B}_1:= a_\mathrm{g}^4 + r^4 + 2 a_\mathrm{g}^2 r^2 \cos 2\theta.
\eqf
Only the one with the minus sign has to be retained since it is finite in the limit $a_\mathrm{g}\rightarrow 0$, contrary to the one with the plus sign which, instead, diverges.
Thus, it is
\begin{align}
\sin^2\theta_\mathrm{BL} \lb{sin2thB}& = \rp{a_\mathrm{g}^2 + r^2 -\sqrt{\mathcal{B}_1}}{2a_\mathrm{g}^2}, \acap
\cos^2\theta_\mathrm{BL} \lb{cos2thB}& = \rp{a_\mathrm{g}^2 - r^2 +\sqrt{\mathcal{B}_1}}{2a_\mathrm{g}^2};
\end{align}
note that both $\sin^2\theta_\mathrm{BL}$ and $\cos^2\theta_\mathrm{BL}$ enter explicitly \rfrs{g00BL}{rho2}.
In the limit $a_\mathrm{g}\rightarrow 0$, \rfrs{sin2thB}{cos2thB} reduce to
\begin{align}
\sin^2\theta_\mathrm{BL} &\rightarrow \sin^2\theta, \acap
\cos^2\theta_\mathrm{BL} &\rightarrow \cos^2\theta.
\end{align}
From \rfrs{sin2thB}{cos2thB}, the differential $d\theta_\mathrm{BL}$, entering explicitly \rfr{ds}, can be obtained in terms of $r$, $\theta$, $dr$, $d\theta$ as
\begin{align}
d\theta_\mathrm{BL} \lb{dthetaB}& = \rp{\ton{-r^2 - a_\mathrm{g}^2 \cos 2\theta + \sqrt{\mathcal{B}_1}}~dr  + a_\mathrm{g}^2 r \sin 2\theta~d\theta}{\sqrt{2\mathcal{B}_1\ton{-r^2 - a_\mathrm{g}^2 \cos 2\theta + \sqrt{\mathcal{B}_1}}}}.
\end{align}

From \rfr{eq2}, calculated with \rfr{sin2thB}, the following two solutions for $r_\mathrm{BL}$, which enters explicitly \rfrs{g00BL}{Delta}, can be obtained in terms of the harmonic coordinates $r$ and $\theta$:
\eqi
r_\mathrm{BL} = \mg \mp\sqrt{\rp{r^2 -a_\mathrm{g}^2 + \sqrt{\mathcal{B}_1}}{2}}.\lb{r2B}
\eqf
Only the solution with the plus sign is acceptable since, in the limit $a_\mathrm{g}\rightarrow 0$, it reduces to 
\eqi
r_\mathrm{BL} \rightarrow \mg + r,
\eqf 
while the one with the minus sign would lead to 
\eqi 
r_\mathrm{BL} \rightarrow \mg - r < 0
\eqf
for $r > \mg$, as in the present case.
Thus, from
\eqi
r_\mathrm{BL} = \mg +\sqrt{\rp{r^2 -a_\mathrm{g}^2 + \sqrt{\mathcal{B}_1}}{2}}\lb{rB}
\eqf
the differential $dr_\mathrm{BL}$, entering explicitly \rfr{ds}, can be obtained in terms of $r$, $\theta$, $dr$, $d\theta$. It reads
\begin{align}
dr_\mathrm{BL} \lb{drB} & = \rp{r\ton{\sqrt{\mathcal{B}_1} + r^2 + a_\mathrm{g}^2\cos 2\theta}~dr - a_\mathrm{g}^2 r^2 \sin2\theta~d\theta}{\sqrt{2\mathcal{B}_1\ton{r^2 -a_\mathrm{g}^2 + \sqrt{\mathcal{B}_1}}}}.
\end{align}

Finally, from \rfr{eq1}, \rfrs{xH}{yH} and \rfrs{xi}{rpm}, the differential $d\phi_\mathrm{BL}$, which enters explicitly \rfr{ds}, contrary to $\phi_\mathrm{BL}$ itself, can be obtained in terms of $r$, $\theta$, $dr$, $d\theta$. It is
\eqi
d\phi_\mathrm{BL} = \rp{\sqrt{2} a_\mathrm{g} \mg^2 r \qua{\ton{r^2 + a_\mathrm{g}^2 \cos 2\theta + \sqrt{\mathcal{B}_1}}~dr + a_\mathrm{g}^2 r \sin 2 \theta~d\theta}}{\sqrt{\mathcal{B}_1} \sqrt{-a_\mathrm{g}^2 + r^2 + \sqrt{\mathcal{B}_1}} \grf{a_\mathrm{g}^4 - \mg^2 r^2 + r^4 + a_\mathrm{g}^2\qua{ -\mg^2 + 2r^2\cos^2\theta} + \ton{a_\mathrm{g}^2 - \mg^2 + r^2} \sqrt{\mathcal{B}_1}}}.\lb{dphiB}
\eqf
By calculating \rfrs{ds}{Delta} with \rfrs{sin2thB}{cos2thB}, \rfrs{dthetaB}{r2B}, and \rfrs{drB}{dphiB} it is possible to express the spacetime metric tensor $g_{\qm\qn}$ of the Kerr metric in terms of the harmonic spherical coordinates. The same task was accomplished in \citet[Equation (15)]{2014GReGr..46.1671J} with the harmonic cartesian coordinates; see also \citet{Ruiz86}. It turns out that it is fully non-diagonal since all its off-diagonal coefficients are nonvanishing. The resulting exact expressions for $g_{\qm\qn}$ are 
%
\begin{align}
g_{00} \lb{g00}& = \rp{\mg^2-\sqrt{\mathcal{B}_1}}{\sqrt{\mathcal{B}_1} + \mg \ton{\mg + \sqrt{2} \sqrt{-\ag^2 + \sqrt{\mathcal{B}_1} + r^2}}},\acap
g_{rr} \nonumber & = \rp{\ton{\sqrt{\mathcal{B}_1} - r^2} \qua{\sqrt{\mathcal{B}_1} + \mg \ton{\mg + \sqrt{2} \sqrt{-\ag^2 + \sqrt{\mathcal{B}_1} + r^2}}}}{2\mathcal{B}_1} +\acap
\nonumber &+\rp{1}{2 \mathcal{B}_1^{3/2} \ton{-\ag^2 + \sqrt{\mathcal{B}_1} + r^2}^{3/2} \ton{\ag^2 + \sqrt{\mathcal{B}_1} - 2 \mg^2 + r^2}}\qua{-\sqrt{2} \ag^2 \mg + \sqrt{2} \mg \ton{\sqrt{\mathcal{B}_1} + r^2} + \right.\acap
\nonumber &\left. + \sqrt{\mathcal{B}_1} \sqrt{-\ag^2 + \sqrt{\mathcal{B}_1} + r^2} + \mg^2 \sqrt{-\ag^2 + \sqrt{\mathcal{B}_1} + r^2}} \grf{\ag^2 \qua{\ton{\ag^2 - \sqrt{\mathcal{B}_1}} \sqrt{\mathcal{B}_1} \ton{\ag^2 + \sqrt{\mathcal{B}_1} - 2 \mg^2} + \right.\right.\acap
\nonumber &\left.\left. + 2 \ton{2 \ag^4 - \mathcal{B}_1 + \sqrt{\mathcal{B}_1} \mg^2} r^2 + 7 \sqrt{\mathcal{B}_1} r^4 + 12 r^6} \cos 2\theta + r^2 \qua{4 r^4 \ton{\sqrt{\mathcal{B}_1} + r^2} + \right.\right.\acap
\nonumber &\left.\left. + \ag^4 \ton{3 \sqrt{\mathcal{B}_1} + 8 r^2} + \ag^4 \ton{\sqrt{\mathcal{B}_1} + 4 r^2} \cos 4\theta}} + \rp{1}{\mathcal{B}_2}\grf{\mg^4 r^2 \ton{\ag^2 - \sqrt{\mathcal{B}_1} + r^2} \qua{4 r^4 \ton{\sqrt{\mathcal{B}_1} + r^2} + \right.\right.\acap
\nonumber &\left.\left. + \ag^4 \ton{3 \sqrt{\mathcal{B}_1} + 8 r^2} + 4 \ag^2 \ton{\ag^4 + 2 \sqrt{\mathcal{B}_1} r^2 + 3 r^4} \cos 2\theta + 
\ag^4 \ton{\sqrt{\mathcal{B}_1} + 4 r^2} \cos 4\theta}}\times\acap
\nonumber &\times \grf{3 \ag^2 + \sqrt{\mathcal{B}_1} + r^2 + \rp{8 \sqrt{2} \mg \ton{\ag^2 + \mg^2}}{\sqrt{-\ag^2 + \sqrt{\mathcal{B}_1} + r^2}} + 
4 \mg \ton{3 \mg + \sqrt{2} \sqrt{-\ag^2 + \sqrt{\mathcal{B}_1} + r^2}} + \right.\acap
\lb{grr} &\left. + \rp{2 \qua{2 \ag^4 + \ag^2 \ton{5 \mg^2 - r^2} + \mg^2 \ton{-\sqrt{\mathcal{B}_1} + 2 \mg^2 + r^2} + \ag^2 r^2 \cos 2\theta}}{-\ag^2 + \sqrt{\mathcal{B}_1} + r^2}},\acap
g_{\theta\theta}\nonumber & = \rp{1}{2 \mathcal{B}_1 \ton{-\ag^2 + \sqrt{\mathcal{B}_1} + r^2}^{3/2} \ton{\ag^2 + \sqrt{\mathcal{B}_1} - 2 \mg^2 + r^2} \ton{\sqrt{\mathcal{B}_1} - r^2 - \ag^2 \cos 2\theta}}\times\acap
\nonumber & \times\ag^4 r^2 \grf{\sqrt{\mathcal{B}_1} \sqrt{-\ag^2 + \sqrt{\mathcal{B}_1} + r^2} + \mg \qua{\mg \sqrt{-\ag^2 + \sqrt{\mathcal{B}_1} + r^2} + \sqrt{2} \ton{-\ag^2 + \sqrt{\mathcal{B}_1} + r^2}}}\times\acap
\nonumber &\times\qua{-\ag^4 + \mathcal{B}_1 + 2 \ag^2 \mg^2 - 2 \sqrt{\mathcal{B}_1} \ton{\mg^2 - 2 r^2} - r^2 \ton{2 \mg^2 + r^2} - 2 \ag^2 r^2 \cos 2\theta} \sin^2\theta 
+ \acap
\nonumber & + \rp{1}{\mathcal{B}_3}\qua{\ag^4 \mg^4 r^4 \ton{\ag^2 - \sqrt{\mathcal{B}_1} + r^2} \grf{\sqrt{-\ag^2 + \sqrt{\mathcal{B}_1} + r^2}\qua{\ag^4 + \mathcal{B}_1 + 4 \mg^4 + 2 \ag^2 \ton{\sqrt{\mathcal{B}_1} - \mg^2} + \right.\right.\right.\acap
\nonumber &\left.\left.\left. + 14 \mg^2 r^2 + r^4 + 2 \sqrt{\mathcal{B}_1} \ton{5 \mg^2 + r^2} + 2 \ag^2 r^2 \cos 2\theta} + \right.\right.\acap
\lb{gthth}&\left.\left. + 4 \sqrt{2} \mg \ton{-\ag^2 + \sqrt{\mathcal{B}_1} + r^2} \ton{\ag^2 + \sqrt{\mathcal{B}_1} + 2 \mg^2 + r^2}}\sin^2 2\theta},\acap
g_{\phi\phi}\nonumber & = \rp{\ag^2 - \sqrt{\mathcal{B}_1} + r^2}{8 \ag^2 \qua{\sqrt{\mathcal{B}_1} + \mg \ton{\mg + \sqrt{2} \sqrt{-\ag^2 + \sqrt{\mathcal{B}_1} + r^2}}}}\grf{\ag^4 + \mathcal{B}_1 + 4 \mg^4 + 14 \mg^2 r^2 + r^4 + \right.\acap
\nonumber &\left. +  4\sqrt{2}\mg\ton{2\mg^2 +  r^2}\sqrt{-\ag^2 + \sqrt{\mathcal{B}_1} + r^2}  + 2 \sqrt{\mathcal{B}_1} \ton{5 \mg^2 + r^2 + 2 \sqrt{2} \mg \sqrt{-\ag^2 + \sqrt{\mathcal{B}_1} + r^2}} + \right.\acap
\lb{gphph} &\left. + 2 \ag^2 \qua{\sqrt{\mathcal{B}_1} + \mg \ton{-\mg + 2 \sqrt{2} \sqrt{-\ag^2 + \sqrt{\mathcal{B}_1} + r^2}}} + 2 \ag^2 r^2 \cos 2\theta},\acap
g_{0r} \lb{g0r}& = \rp{\mg^3 r \ton{\ag^2 - \sqrt{\mathcal{B}_1} + r^2} \ton{2 \mg + \sqrt{2} \sqrt{-\ag^2 + \sqrt{\mathcal{B}_1} + r^2}} \ton{\sqrt{\mathcal{B}_1} + r^2 + \ag^2 \cos 2\theta}}{\mathcal{B}_4},\acap
g_{0\theta}\lb{g0th} & = -\rp{\ag^2 \mg^3 r^2 \ton{\ag^2 - \sqrt{\mathcal{B}_1} + r^2} \ton{2 \mg + \sqrt{2} \sqrt{-\ag^2 + \sqrt{\mathcal{B}_1} + r^2}} \sin 2\theta}{\mathcal{B}_4},\acap
g_{0\phi}\lb{g0ph} & = -\rp{\mg \ton{\ag^2 - \sqrt{\mathcal{B}_1} + r^2} \ton{2 \mg + \sqrt{2} \sqrt{-\ag^2 + \sqrt{\mathcal{B}_1} + r^2}}}{2 \ag \qua{\sqrt{\mathcal{B}_1} + \mg \ton{\mg + \sqrt{2} \sqrt{-\ag^2 + \sqrt{\mathcal{B}_1} + r^2}}}},\acap
g_{r\theta}\nonumber & = -\rp{\ag^2 \mg^2 r \qua{\sqrt{\mathcal{B}_1} + \mg \ton{\mg + \sqrt{2} \sqrt{-\ag^2 + \sqrt{\mathcal{B}_1} + r^2}}} \sin 2\theta}{\mathcal{B}_1 \ton{\ag^2 + \sqrt{\mathcal{B}_1} - 2 \mg^2 + r^2}}-\acap
\nonumber &-\rp{\ag^2 \mg^4 r^3 \ton{\ag^2 - \sqrt{\mathcal{B}_1} + r^2} \qua{\ag^4 + \sqrt{\mathcal{B}_1} r^2 + r^4 + \ag^2 \ton{\sqrt{\mathcal{B}_1} + 2 r^2} \cos 2\theta}}{\mathcal{B}_5}\times\acap
\nonumber&\times\grf{\qua{\ag^2 + \rp{1}{4} \ton{2 \mg + \sqrt{2} \sqrt{-\ag^2 + \sqrt{\mathcal{B}_1} + r^2}}^2}^2 + \right.\acap
\lb{grth}& + \left. \rp{1}{2} \qua{\ag^2 \ton{\mg^2 - r^2} + \mg^2 \ton{-\sqrt{\mathcal{B}_1} + r^2} + \ag^2 r^2 \cos 2\theta}} \sin 2\theta,\acap
g_{r\phi}\nonumber & = -\rp{\mg^2 r \ton{\ag^2 - \sqrt{\mathcal{B}_1} + r^2} \ton{\sqrt{\mathcal{B}_1} + r^2 + \ag^2 \cos 2\theta}}{\ag H}\times\acap
\lb{grph} &\times \grf{\qua{\ag^2 + \rp{1}{4} \ton{2 \mg + \sqrt{2} \sqrt{-\ag^2 + \sqrt{\mathcal{B}_1} + r^2}}^2}^2 + \rp{1}{2} \qua{\ag^2 \ton{\mg^2 - r^2} + \mg^2 \ton{-\sqrt{\mathcal{B}_1} + r^2} + \ag^2 r^2 \cos 2\theta}},\acap
g_{\theta\phi}\nonumber & = \rp{\ag \mg^2 r^2 \ton{\ag^2 - \sqrt{\mathcal{B}_1} + r^2}}{\mathcal{B}_4}\grf{\qua{\ag^2 + \rp{1}{4} \ton{2 \mg + \sqrt{2} \sqrt{-\ag^2 + \sqrt{\mathcal{B}_1} + r^2}}^2}^2 + \right.\acap
\lb{gthph}&\left. + \rp{1}{2} \qua{\ag^2 \ton{\mg^2 - r^2} + \mg^2 \ton{-\sqrt{\mathcal{B}_1} + r^2} + \ag^2 r^2 \cos 2\theta}} \sin 2\theta.
\end{align}
The auxiliary functions $\mathcal{B}_2,\mathcal{B}_3,\mathcal{B}_4,\mathcal{B}_5$ of $r,\theta$ and $\ag,\mg$ which were introduced in \rfrs{g00}{gthph} are
\begin{align}
\mathcal{B}_2\lb{L} \nonumber & := 8 \mathcal{B}_1^{3/2}  \qua{\sqrt{\mathcal{B}_1} + \mg \ton{\mg + \sqrt{2} \sqrt{-\ag^2 + \sqrt{\mathcal{B}_1} + r^2}}}\times\acap
&\times \qua{\ton{\ag^2 + \sqrt{\mathcal{B}_1}} \ton{\ag^2 - \mg^2} + \ton{\ag^2 + \sqrt{\mathcal{B}_1} - \mg^2} r^2 + r^4 + \ag^2 r^2 \cos 2\theta}^2,\acap
\mathcal{B}_3 \nonumber &:= 4 \ton{-\ag^2 + \sqrt{\mathcal{B}_1} + r^2}^{3/2} \qua{\mathcal{B}_1^{3/2} + \mathcal{B}_1 \mg \ton{\mg + \sqrt{2} \sqrt{-\ag^2 + \sqrt{\mathcal{B}_1} + r^2}}}\times\acap
\lb{K}&\times \qua{\ton{\ag^2 + \sqrt{\mathcal{B}_1}} \ton{\ag^2 - \mg^2} + \ton{\ag^2 + \sqrt{\mathcal{B}_1} - \mg^2} r^2 + r^4 + \ag^2 r^2 \cos 2\theta}^2, \acap
\mathcal{B}_4 \nonumber &:= \sqrt{2\mathcal{B}_1} \sqrt{-\ag^2 + \sqrt{\mathcal{B}_1} + r^2} \qua{\sqrt{\mathcal{B}_1} + \mg \ton{\mg + \sqrt{2} \sqrt{-\ag^2 + \sqrt{\mathcal{B}_1} + r^2}}}\times\acap
\lb{H} &\times \qua{\ton{\ag^2 + \sqrt{\mathcal{B}_1}} \ton{\ag^2 - \mg^2} + \ton{\ag^2 + \sqrt{\mathcal{B}_1} - \mg^2} r^2 + r^4 + \ag^2 r^2 \cos 2\theta},\acap
\mathcal{B}_5\nonumber & := \mathcal{B}_1^{3/2} \ton{-\ag^2 + \sqrt{\mathcal{B}_1} + r^2} \ton{\sqrt{\mathcal{B}_1} + \mg^2 + \sqrt{2} \mg \sqrt{-\ag^2 + \sqrt{\mathcal{B}_1} + r^2}}\times\acap
\lb{N} & \times \qua{\ag^4 - \mg^2 r^2 + r^4 + \ag^2 \ton{-\mg^2 + r^2} + \sqrt{\mathcal{B}_1} \ton{\ag^2 - \mg^2 + r^2} + \ag^2 r^2 \cos 2\theta}^2.
\end{align}
By expanding \rfrs{g00}{gthph} in powers of $a_\mathrm{g}$ and $\mg$, it is possible to obtain approximate expressions for them in agreement with those of Equations (80)-(83) in \citet{2008PhRvD..77j4001H}. In the following, the power expansion will be pushed beyond the order strictly required to match the results by \citet{2008PhRvD..77j4001H} in order to fully capture  the accelerations of the order of $\mathcal{O}\ton{1/c^4}$, as shown in Section \ref{sec:3}.
\appendixsection{}\lb{app:B}
In the literature, the Kerr metric is found generally written in a coordinate system whose $z$ axis is aligned with the hole's spin axis $\bds{\hat{k}}$, namely
\eqi
\bds{\hat{k}}=\grf{0,0,1}.\lb{k01}
\eqf 
If, on the one hand, this leads to more manageable calculations, on the other hand, the resulting expressions lack the generality required in contexts where the spatial orientation of the black hole's spin is often poorly constrained, if any. Similar considerations also apply in the case where the primary is a rotating material body such as a neutron star, a planet, etc.

Here, a general method to write the components of any acceleration depending on the primary's spin axis, assumed as in \rfr{k01},  in a compact vector form valid for an arbitrary orientation of $\bds{\hat{k}}$ is presented.

In the following, the unprimed quantities refer to a coordinate system $K$ where \rfr{k01} holds.
First, $\bds{\hat{k}}$ has to be parameterized in terms of two spherical angles characterizing its desired generic direction as, say, 
\eqi
\bds{\hat{k}} =\grf{\sin\beta\sin\gamma,\sin\beta\cos\gamma,\cos\beta}.\lb{kap}
\eqf
Then, a coordinate transformation rotating  $K$ to a new primed coordinate system $K^{'}$ whose $z^{'}$ axis is aligned with \rfr{kap} is considered. This task is implemented by the rotation matrix
\eqi
\mathcal{R}=\left(
  \begin{array}{ccc}
  \cos\alpha \cos\gamma - \cos\beta \sin\alpha \sin\gamma   & \cos\gamma \sin\alpha + \cos\alpha \cos\beta \sin\gamma & \sin\beta\sin\gamma \\
  -\cos\beta \cos\gamma \sin\alpha - \cos\alpha \sin\gamma   & \cos\alpha \cos\beta \cos\gamma - \sin\alpha \sin\gamma & \sin\beta\cos\gamma \\
  \sin\alpha \sin\beta   & -\cos\alpha \sin\beta & \cos\beta \\
  \end{array}
\right).\lb{kazzo}
\eqf
The angle $\alpha$ in \rfr{kazzo} is an unconstrained degree of freedom representing a rotation in the $\grf{x^{'},y^{'}}$ plane about \rfr{kap}; it will finally not enter the transformed accelerations. The inverse matrix $\mathcal{R}^{-1}$ of \rfr{kazzo}, bringing \rfr{kap} back to the $z$ axis, has now to be calculated and applied to the position and velocity vectors intended as referred to $K^{'}$. The resulting expressions for their rotated back components $x_1,x_2,x_3,v_1,v_2,v_3$ are just those entering the accelerations obtainable from the standard expressions of the Kerr metric valid for the case of \rfr{k01}. Finally, the acceleration in $K$, now written in terms of  $x_1,x_2,x_3,v_1,v_2,v_3$ as functions of  $x_1^{'},x_2^{'},x_3^{'},v_1^{'},v_2^{'},v_3^{'}$, has to be rotated by \rfr{kazzo}; at this stage, the primes can be discarded. The resulting formulas, which turn out to be independent of $\alpha$, give the desired components of the acceleration for an arbitrary orientation of the primary's spin axis. Finally, 
by using
\begin{align}
\beta & =\arccos\ton{{\hat{k}}_z},\acap
\gamma &=\arctan\ton{\rp{{\hat{k}}_x}{{\hat{k}}_y}},
\end{align}
a compact vector form where $\bds{\hat{k}}$ enters can be obtained.
\section*{Data availability}
No new data were generated or analysed in support of this research.
\section*{Conflict of interest statement}
I declare no conflicts of interest.
\section*{Funding}
This research received no external funding.
\section*{Acknowledgements}
I thank some anonymous referees for their constructive and useful comments.

\bibliography{Megabib}{}
\end{document}